\documentclass[11pt,a4paper]{article}

\usepackage{ascmac}
\usepackage{amsmath}
\usepackage{amssymb}
\usepackage{bm}
\usepackage[footnotesize,bf]{caption}
\usepackage{fullpage}
\usepackage{color}
\usepackage{float}
\usepackage{graphicx}
\usepackage[]{natbib}
\usepackage{subfigure}
\usepackage{txfonts}
\usepackage{threeparttable}
\usepackage{wrapfig}
\usepackage[]{multicol}
\usepackage{wrapfig}
\usepackage{authblk}

\usepackage{floatflt}

\title{\large Complete list of the ASTRO-H Science Working Group}
\date{\vspace{-0.5cm}}
\newcommand{\MakeWhitePaperTitle}{
	\begin{center}
		\begin{figure}
			\vspace{1cm}
			\begin{center}
				\includegraphics[width=0.2\hsize]{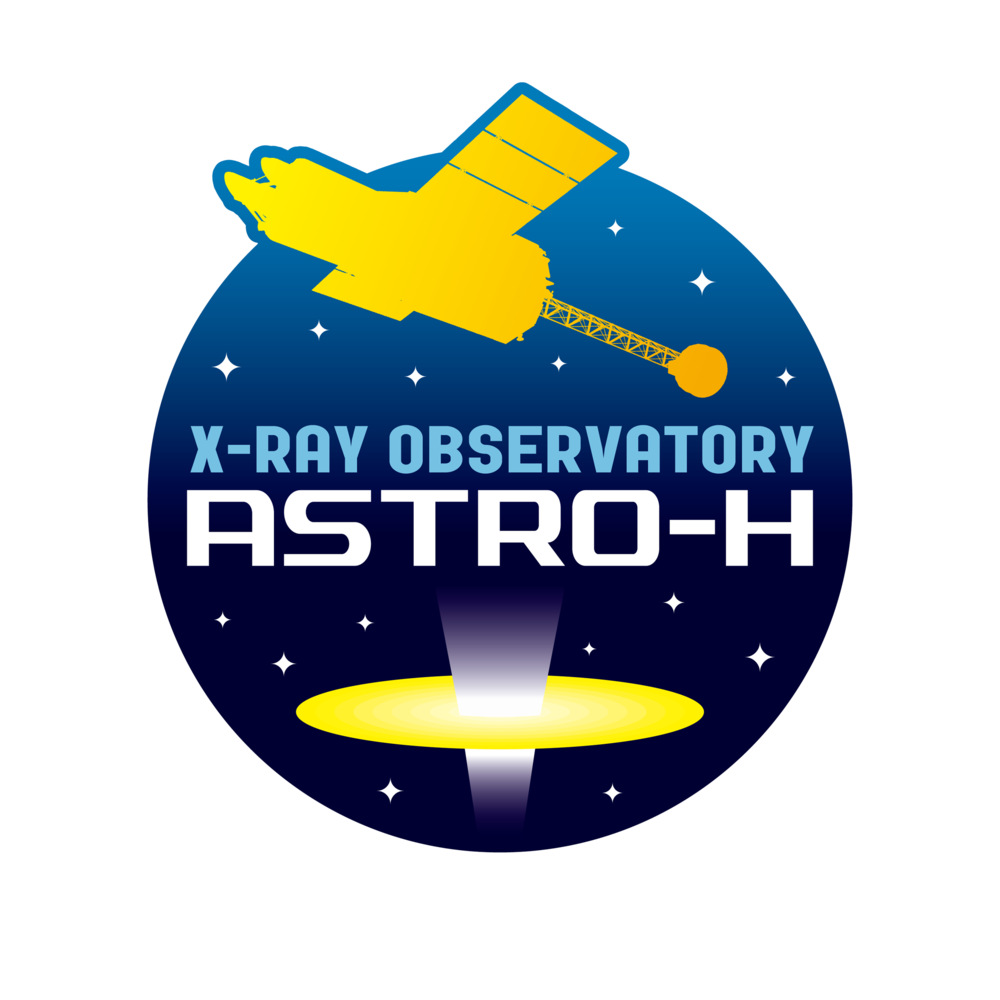}
			\end{center}
		\end{figure}
		\vspace{1cm}
		{\LARGE
		ASTRO-H Space X-ray Observatory\\
		White Paper\\
		}
		\vspace{5mm}
		{\large
		\WhitePaperTitle\\
		}
		\vspace{1cm}
		{
		\WhitePaperAuthors\\
		on behalf of the ASTRO-H Science Working Group
		}
	\end{center}
}

\usepackage{authblk}
\author[a]{Tadayuki~Takahashi}
\author[a]{Kazuhisa~Mitsuda}
\author[b]{Richard~Kelley}
\author[c]{Felix~Aharonian}
\author[d]{Hiroki~Akamatsu}
\author[e]{Fumie~Akimoto}
\author[f]{Steve~Allen}
\author[g]{Naohisa~Anabuki}
\author[b]{Lorella~Angelini}
\author[h]{Keith~Arnaud}
% ...
\affil[a]{Institute of Space and Astronautical Science (ISAS), Japan Aerospace Exploration Agency (JAXA), Kanagawa 252-5210, Japan}
\affil[b]{NASA/Goddard Space Flight Center, MD 20771, USA}
\affil[c]{Astronomy and Astrophysics Section, Dublin Institute for Advanced Studies, Dublin 2, Ireland}
\affil[d]{SRON Netherlands Institute for Space Research, Utrecht, The Netherlands}
\affil[e]{Department of Physics, Nagoya University, Aichi 338-8570, Japan}
\affil[f]{Kavli Institute for Particle Astrophysics and Cosmology, Stanford University, CA 94305, USA}
\affil[g]{Department of Earth and Space Science, Osaka University, Osaka 560-0043, Japan}
\affil[h]{Department of Astronomy, University of Maryland, MD 20742, USA}
% ...

\def\cm{{\rm\thinspace cm}}

\def\erg{{\rm\thinspace erg}}

\def\keV{{\rm\thinspace keV}}

\def\s{{\rm\thinspace s}}

\def\cmps{\hbox{$\cm\s^{-1}\,$}}

\def\ergpcmsqps{\hbox{$\erg\cm^{-2}\s^{-1}\,$}}

\def\rg{\hbox{$r_{\rm g}$}}

\begin{document}

\newcommand{\WhitePaperTitle}{AGN Reflection}
\newcommand{\WhitePaperAuthors}{
C.~Reynolds (University of Maryland),
Y.~Ueda (Kyoto University),
H.~Awaki (Ehime University),\\
L.~Gallo (Saint Mary's University),
P.~Gandhi (University of Durham\footnote{Also at University of Southampton}), 
\\
Y.~Haba (Aichi University of Education),
T.~Kawamuro (Kyoto University),
S.~LaMassa (Yale University), 
A.~Lohfink (University of Maryland\footnote{Also at University of Cambridge}),
C.~Ricci (Kyoto University),
F.~Tazaki (Kyoto University\footnote{Also at National Astronomical Observatory of Japan}),\\
and A.~Zoghbi (University of Maryland\footnote{Also at University of Michigan})
}
\MakeWhitePaperTitle

\begin{figure}[h]
\centerline{
\includegraphics[width=0.8\textwidth]{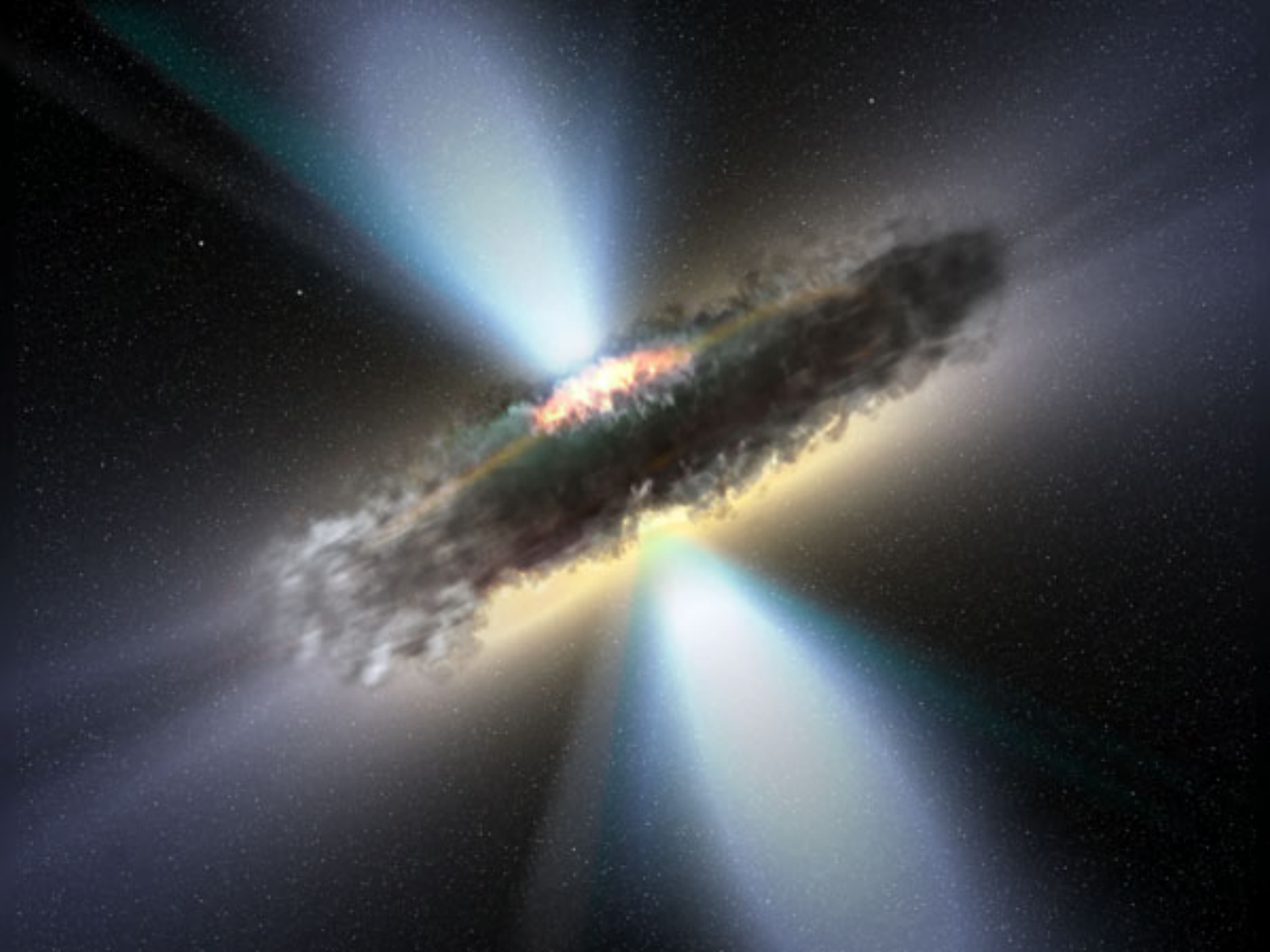}
}
\label{fig:title}
\end{figure}
Figure Credit: ESA / V. Beckmann (NASA)

\newpage

\begin{abstract}
X-ray observations provide a powerful tool to probe the central engines of active galactic nuclei (AGN).  A hard X-ray continuum is produced from deep within the accretion flow onto the supermassive black hole, and all optically thick structures in the AGN (the dusty torus of AGN unification schemes, broad emission line clouds, and the black hole accretion disk) ``light up'' in response to irradiation by this continuum.  This White Paper describes the prospects for probing AGN physics using observations of these X-ray reflection signatures.  High-resolution SXS spectroscopy of the resulting fluorescent iron line in type-2 AGN will give us an unprecedented view of the obscuring torus, allowing us to assess its dynamics (through line broadening) and geometry (through the line profile as well as observations of the ``Compton shoulder'').  The broad-band view obtained by combining all of the {\it ASTRO-H} instruments will fully characterize the shape of the underlying continuum (which may be heavily absorbed) and reflection/scattering, providing crucial constraints on models for the Cosmic X-ray Background with a subsequent impact on understanding of supermassive black hole evolution.  {\it ASTRO-H} will also permit the relativistically broadened reflection spectrum from the inner accretion disk to be robustly studied, even in complex systems with, for example, warm absorption and composite soft excesses.  Finally, the HXI will allow the detection and study of reverberation delays between the continuum and the Compton reflection hump from the inner disk.  
\end{abstract}

\maketitle
\clearpage

\tableofcontents
\clearpage

\section{General Introduction}

After over four decades of study, active galactic nuclei (AGN) continue to be of great interest to astrophysicists.  With the realization that supermassive black holes (SMBHs) are ubiquitous, being present at the center of essentially every galaxy, a tremendous amount of attention is focused on the co-evolution of SMBHs and galaxies.  The tight correlation between the mass of the SMBH in a galactic center and the velocity dispersion of the stellar budge \citep{gebhardt:00a,ferrarese:00a} found in the local universe strongly suggests that SMBHs and host galaxies co-evolved.  In the extreme (probably realized in massive galaxies), a SMBH can dramatically suppress baryonic cooling and star formation in the galaxy, all-but truncating any further growth of the galaxy.  AGN are also interesting in their own right, providing one of the most accessible laboratories for studying the extremes of physics found close to the horizon of a black hole (BH).

\begin{figure}[h]
\centerline{
\includegraphics[width=0.7\textwidth]{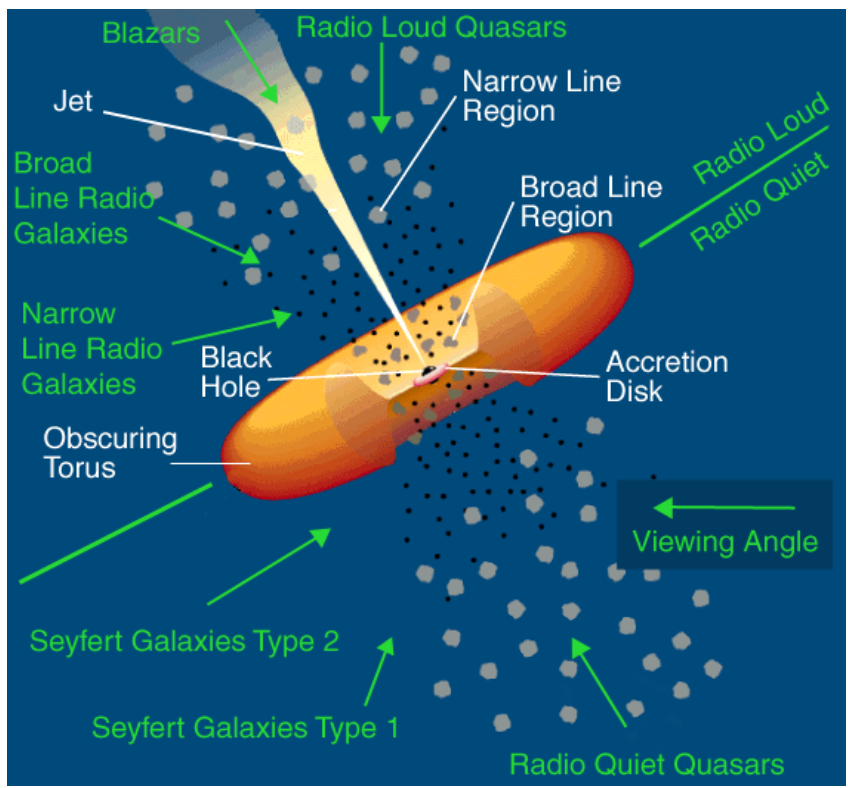}
}
\caption{The anatomy of an AGN as according to the standard unified model.   Figure from \cite{urry:95a}}
\label{fig:unified_agn}
\end{figure}

Any study of AGN must start by characterizing their ``engineering blueprint", i.e., the properties of the SMBH (mass and spin), the distribution of circumnuclear matter, the (inward and outward) flow of matter, and the forms taken by the liberated energy as it leaves the AGN.  Our current understanding is summarized in Figure~\ref{fig:unified_agn}.  At its heart, an AGN consists of a SMBH fed by an accretion disk.  Powerful winds can be produced from the accretion disk which probably produce the optical broad line region (BLR) with characteristic velocities of 2,000--20,000\,${\rm km}\,{\rm s}^{-1}$.  These broad optical lines are the defining characteristic of the classical unabsorbed (type-1) AGN.  Further out, we know that many AGN possess a cold and dusty structure (the ``dusty torus") that can, depending upon viewing angle, obscure the BLR and the central accretion disk, leading to a classification as an obscured (type-2) AGN.  The nature (and even the location) of this important structure is unclear.  The torus may be the reservoir for the accretion disk, being fed by cold gas flows from the ISM of the galaxy --- in other words, it is gas on the way in.   Alternatively, it may be the colder/outer regions of the disk wind --- gas on the way out.  Both possibilities may be realized across the AGN population.  In either case, ``the torus'' is the interface of the AGN with the rest of the galaxy, and understanding its basic nature
is crucial if we are to understand the AGN-galaxy link.  

\begin{figure}[h]
\centerline{
\includegraphics[width=0.7\textwidth]{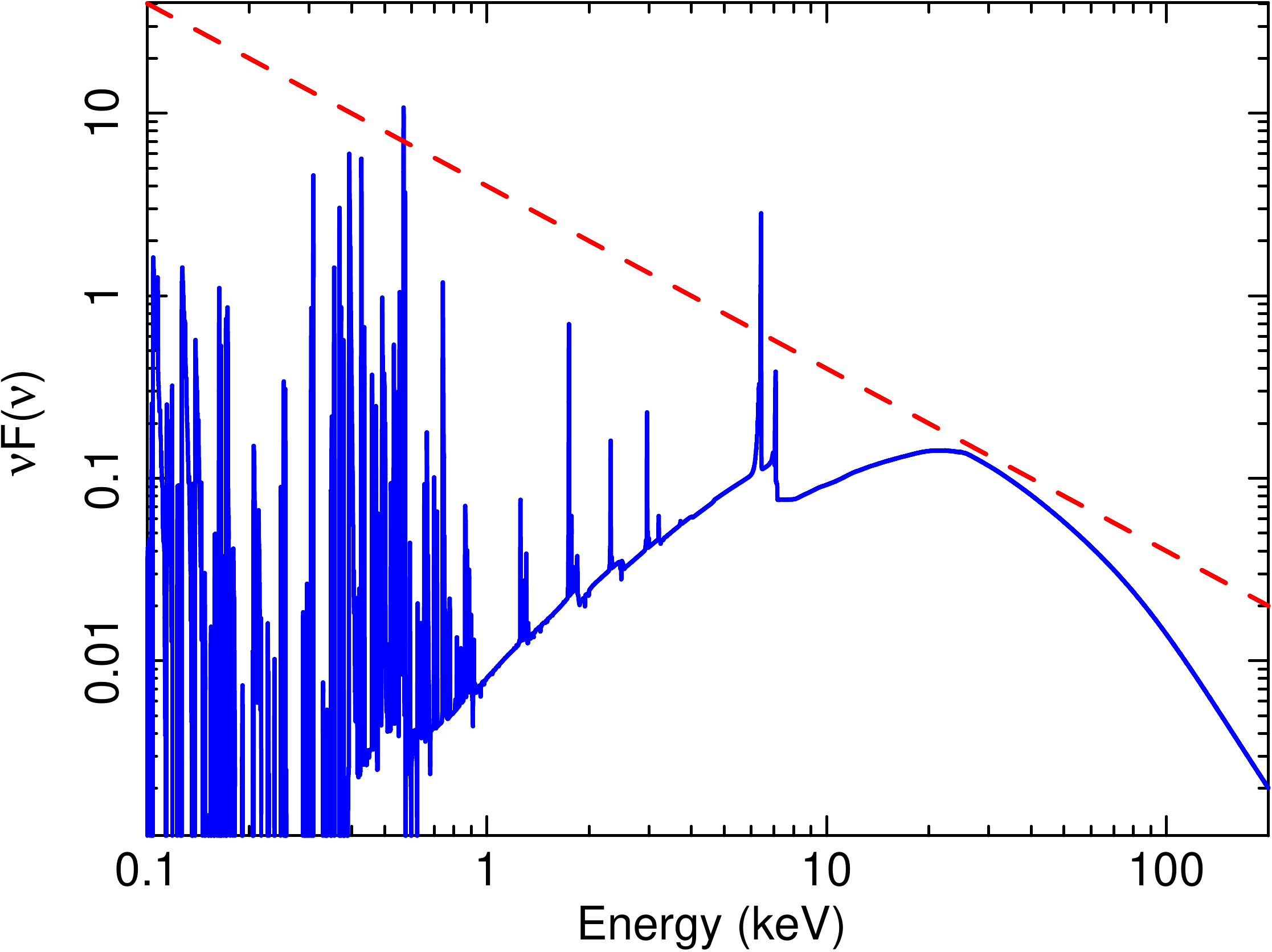}
}
\caption{Example of an X-ray reflection spectrum, assuming a slab of matter irradiated by a power-law spectrum with a photon index $\Gamma=2$ (red dashed line).  The surface of the slab is assumed to have an ionization parameter of $\xi=1\,{\rm erg}\,{\rm cm}\,{\rm s}^{-1}$.  Figure produced using the {\tt xillver} code of \cite{garcia:13a}}
\label{fig:reflection}
\end{figure}

X-ray spectroscopy is a powerful tool for determining the engineering blueprint of an AGN.  X-rays are produced deep within the heart of the AGN, within a few gravitational radii ($r_g\equiv GM/c^2$, where $M$ is the mass of the SMBH) of the black hole.  Any optically-thick structures in the AGN (the accretion disk, BLR clouds, and the torus) will then be irradiated by the X-ray continuum and respond by producing an ``X-ray reflection spectrum''.  An example of an X-ray reflection spectrum is shown in Figure~\ref{fig:reflection}.  The classic signatures of X-ray reflection are a strong iron-K$\alpha$ emission line (at energies 6.4--6.97\,keV depending upon the ionization state of the reflector) and a broad hard X-ray hump peaking at 20\,keV where Compton reflection dominates over photoelectric absorption (at lower energies) and Compton recoil losses (at higher energies).  If the reflector has a moderate-to-high ionization parameter, as is the case for the inner accretion disk, the reflection spectrum also possesses a forest of soft X-ray radiative-recombination features.

The iron line is a crucial diagnostic of AGN structure; iron is an abundant element and possess strong emission/absorption features that are relatively isolated from other spectral complexities.   The iron line is to AGN/X-ray astronomers what the CO rotational lines are to star-formation/mm-wave astronomers.  Detailed studies of reflection spectra allow us to study the distribution and kinematics of gas flowing in, out and around the SMBH; with this tool,
we can start to determine the blueprint of the AGN central engine.

This White Paper (WP) discusses the impact of {\it ASTRO-H} on studies of AGN structure using X-ray reflection.  The WP is divided into two parts.  We start (Sections~\ref{sec:distant} and \ref{sec:comptonthick}) by discussing the ability of {\it ASTRO-H} to map out the structures that are distant from the SMBH.   Section~\ref{sec:distant} discusses the impact of {\it ASTRO-H} on studies of photoionized winds and the torus in general type-2 AGN, whereas Section~\ref{sec:comptonthick} focuses on the particularly interesting case of Compton-Thick AGN.   The superior spectral resolution of the {\it ASTRO-H}/SXS and the superior bandbass achieved by combining SXS and HXI data will transform our knowledge of these systems.  As already mentioned, an understanding of AGN winds and the torus is crucial if we wish to truly understand how the AGN is fueled and how it feeds back on its host galaxy.  The second part of this WP (Section~\ref{sec:relrefl}) discusses the ability of {\it ASTRO-H} to study X-ray reflection from the inner accretion disk, where relativistic Doppler and gravitational redshifts can strongly skew the observed spectrum.  Reflection from the inner disk is a broad-band phenomenon, resulting in a soft excess (as the radiative recombination lines are smeared into a pseudo-continuum), the broadened iron line, and the Compton reflection hump.  {\it ASTRO-H} will be the first observatory capable of simultaneously observing and characterizing in detail all three aspects of disk reflection.  This energy reach represents an important advance.

\section{Mapping the structure of obscured AGN}\label{sec:distant}

\subsection{The X-ray View of Type-2 AGN}

The basic nature of dusty tori in AGN remains unclear.  We are ignorant about many of their most basic characteristics, including their location, size, total mass, clumpiness, dynamics, and relation to the accretion disk and BLR.  Even the ``definition'' of the torus is ambiguous, and it is not
entirely clear whether it is distinct structure or just a smooth continuation from the BLR.  A particularly curious fact is that they must typically subtend a large solid angle as seen from the SMBH in order to produce the observed fraction of type-2 AGN.    Theoretical studies suggest that it is not easy to stably sustain such a cold but large scale-height ($h/r \sim 1$) structure. This implies that the torus is not just static structure but may be a sequence of dynamic phenomena like nuclear starburst \citep[e.g.,][]{hopkins12,wada12}. The total mass of a torus may be an important parameter reflecting the evolutional stage of the galaxy \citep{kawakatu08}.

High quality broad-band X-ray spectra give unique insights into the anatomy of an AGN. Hard X-rays have strong penetrating power against obscuration, enabling us to study the properties of the obscuring matter and surrounding gas. Type-2 AGN are particularly well suited for a study of AGN structure; (1) information on the line-of-sight material can be obtained through X-ray absorption, (2) photoionized gas located outside the torus (in the narrow line region; NLR) can be observed in emission and scattering lines that gives definite probes on its physical state, (3) the equivalent widths of iron-K lines from the obscuring torus become large thanks to the attenuated continuum level and hence are more readily characterized, and (4) the high inclination angles make it easier to measure Keplerian motion of line emitting matter.

The typical X-ray spectrum of a type-2 AGN is characterized by the combination of a heavily absorbed primary power-law, Compton reflection with fluorescence lines from the cold torus (with additional reflection possible from the BLR and accretion disk), and soft X-ray emission dominated by the emission lines from highly ionized gas \citep{turner97}. The soft emission lines can provide diagnostic tools to reveal the temperature, density, and ionization state of the gas surrounding the supermassive black hole.  

\begin{figure*}[htbp]
\begin{center}
\resizebox{0.48\hsize}{!}{\includegraphics[angle=-90]{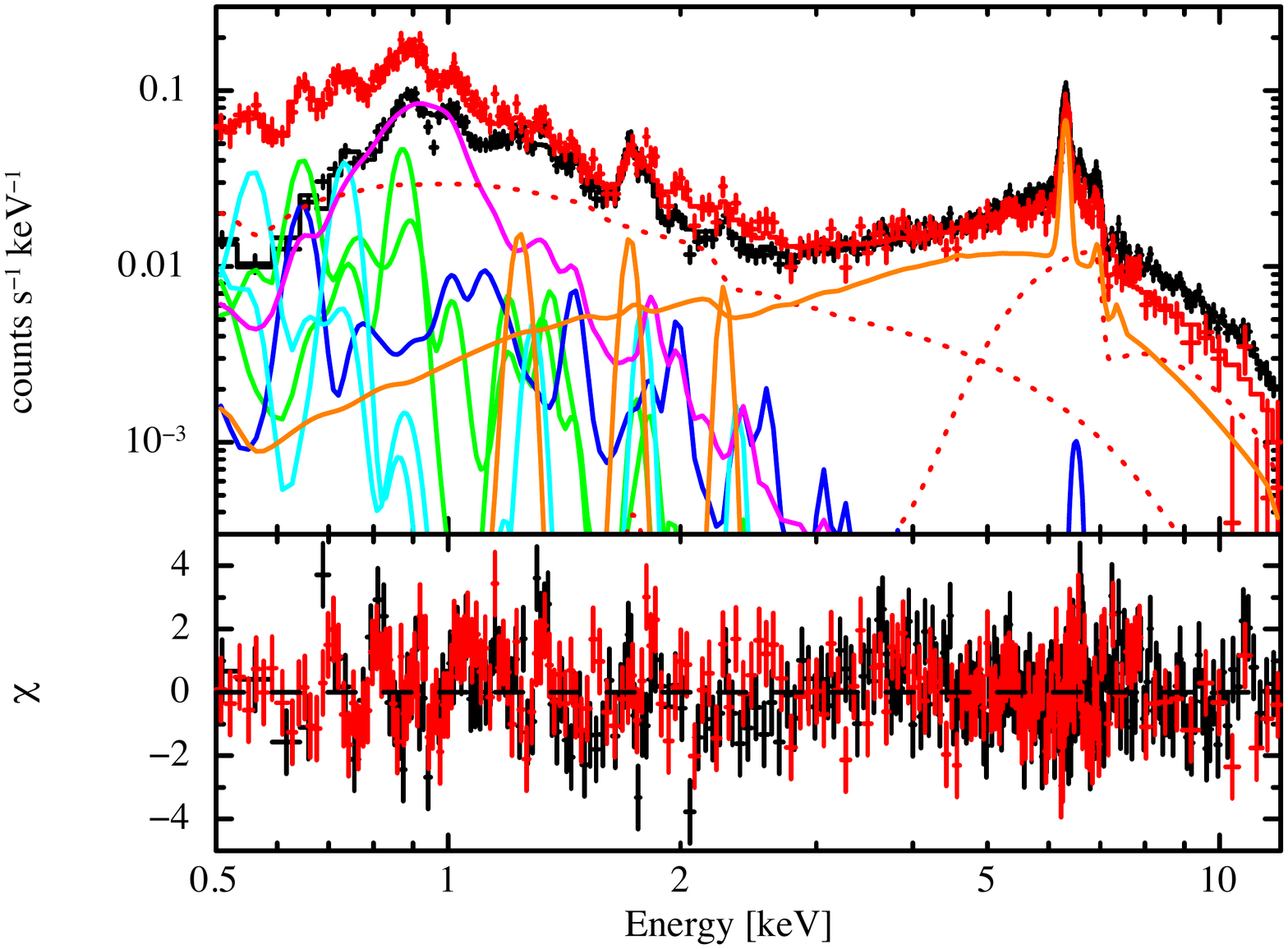}}
\,\,
\resizebox{0.47\hsize}{!}{\includegraphics[angle=-90]{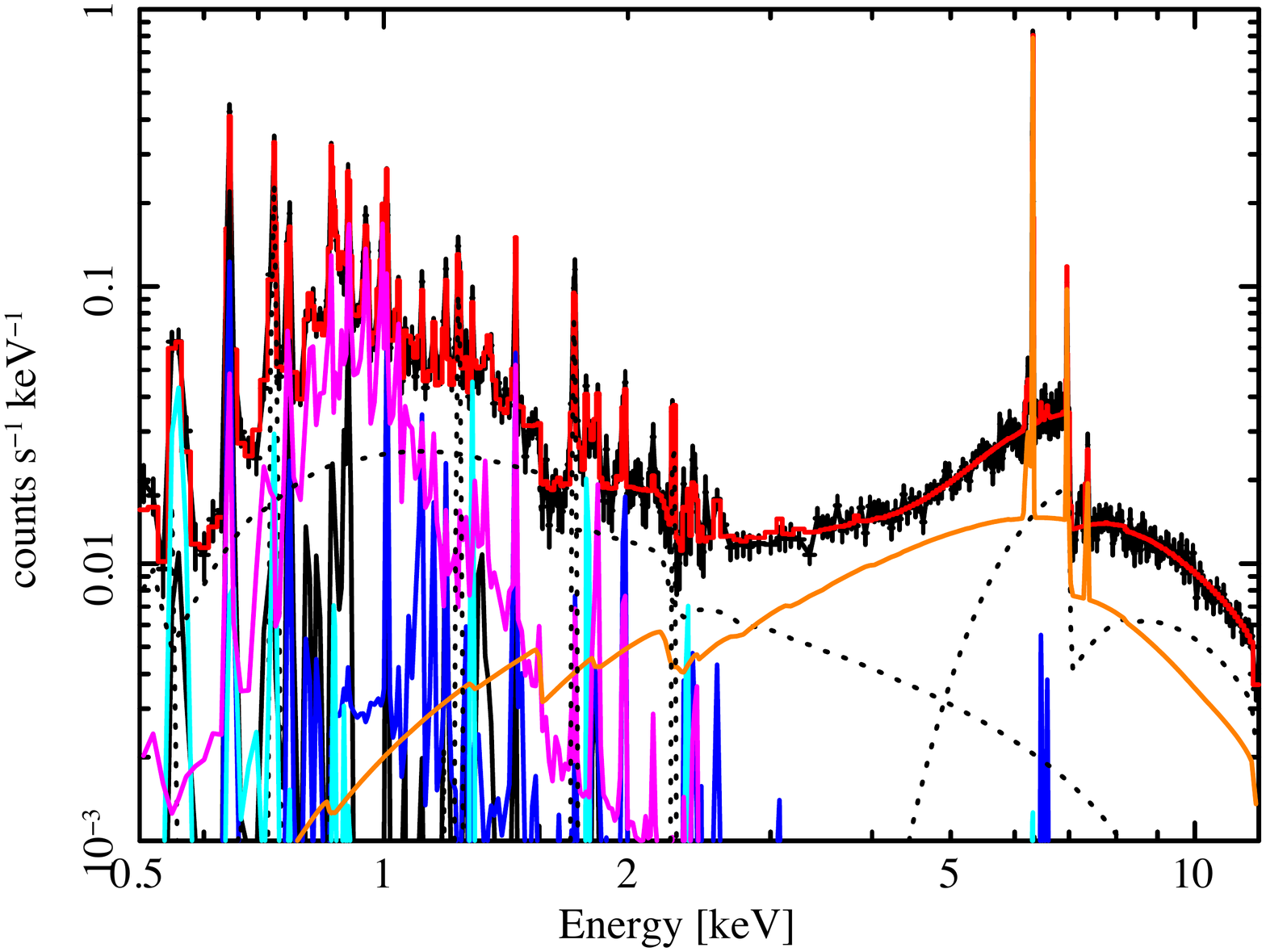}}
\end{center}
\caption{{\it Left panel : } The spectrum of Mrk~3 as observed by the {\it Suzaku}/XIS in 2005.  The continuum (direct/absorbed an scattered/unabsorbed) is shown as dotted orange line.  A strong cold reflection component is marked as a solid orange line.  The complex soft spectrum is described as coming from three different photoionized plasmas (light blue; $\log \xi$ $=$ 0.01, $N_{\rm H}$ $=$1.0$\times$10$^{20}$ cm$^{-2}$, green; $\log \xi$ $=$ 1.8, $N_{\rm H}$
$=$ 1.4$\times$10$^{21}$ cm$^{-2}$, blue; $\log \xi$ $=$ 2.9, $N_{\rm H}$ $=$ 2.0$\times$10$^{22}$ cm$^{-2}$) and an optically-thin thermal plasma (magenta; \texttt{apec} in XSPEC) with temperature $kT$ of 1.0 keV.   {\it Right panel : }  The simulated SXS spectrum of Mrk~3 for a 200\,ksec exposure based on the spectral model for Mrk~3.  We employ the redistribution matrix function (RMF) with 5\,eV resolution, the current best-estimated (file {\tt ah\_sxs\_5ev\_basefilt\_20100712.rmf}), and the ancillary response file (ARF) for the point source (file {\tt sxt-s\_120210\_ts02um\_of\_intallpxl.arf}).}
\label{fig:mrk3_spec}
\end{figure*}
As an illustration, Figure~\ref{fig:mrk3_spec} (left panel) shows the {\it Suzaku} spectrum of the prototypical Seyfert-2 galaxy Mrk~3 ($z=0.0135$), one of the brightest objects among the class \citep{cappi99}.   Despite having an absorption column that is marginally Compton-Thick, we can see still see the direct (absorbed) power-law continuum at high-energies.  We also see that some fraction of this continuum is scattered around the the torus.   A cold X-ray reflection component from, presumably, Compton-Thick regions of the torus out of our line of sight produces a strong 6.4\,keV iron fluorescence line and contributes significantly to the continuum above 5\,keV.  The soft-band is very complex \citep{sako00,poupag05,awaki08}, requiring photoionized emitters with at least three different ionization parameters as well as a collisionally ionized thermal plasma component (see figure caption).  

\subsection{{\it ASTRO-H}/SXS studies of type-2 AGN}

The superior resolution of the {\it ASTRO-H}/SXS will permit a giant leap forward in the study of these line-rich AGN.  For example, Figure~\ref{fig:mrk3_spec} (right panel) shows a simulation of a 200\,ks SXS observation of Mrk~3.  Comparing with the {\it Suzaku}/XIS spectrum, we can see that the forrest of soft X-ray emission lines is resolved, allowing us to employ the full machinery of  line ratio diagnostics for density, temperature, and excitation mechanism.  The K$\alpha$-triplet lines from helium-like ions are particularly useful \citep{gabjor69,pordub00}.  While these soft X-ray lines have already been seen at such resolutions by the grating instruments on {\it Chandra} and {\it XMM-Newton},  {\it ASTRO-H}/SXS provides significantly more sensitivity at soft energies (permitting the full spectral resolution to be realized in realistic exposure times) and, simultaneously, unprecedented resolution of the iron line complex.  

The ability to resolve and characterize the velocity profile of the fluorescent iron line will permit the first study of the dynamics of the X-ray reflecting torus.     At the same time, the resolution of the SXS will greatly facilitate our ability to centroid the K$\alpha$ line as well as detect the K$\beta$ line --- both of these observables are sensitive to the ionization state of the gas.  Putting this information together will permit us to determine the location ($r$), size ($\Delta r$) and density ($n$) of the torus material --- the velocity broadening $\Delta v$ gives us the location from Keplerian arguments ($\Delta v\approx\sqrt{GM/r}$), the dominant ionization states of iron tell us the ionization parameter ($\xi=L_i/nr^2$, where $L_i$ is the ionizing luminosity) and the combination of these two yields the density.  Comparing the density with the line of sight column density (which can be measured from the continuum absorption, provided that the AGN is Compton-thin) gives the size of the torus $\Delta r$.

While detailed dynamical and ionization models will be needed to interpret the real results, we can obtain a first look at the impact of {\it ASTRO-H}/SXS results by assuming that the torus has a single ionization parameter and the form of a disk-like structure in Keplerian motion about the SMBH.  For this exercise, we use the XSTAR photoionization code to model the scattered and emitted spectrum of a torus with a density of $n=10^{13}$ cm$^{-3}$, a column density of $N_{\rm H} = 10^{23}$cm$^{-2}$ irradiated by a Seyfert-like X-ray source (luminosity of $10^{44}$ erg s$^{-1}$ (0.01--10 keV) and a power-law spectrum of $\Gamma=2.0$).  We adopt three different distances for the torus, 1 pc, 0.1 pc, and 0.01 pc, which we designate the ``far torus'', ``near torus'', and ``BLR'' cases, respectively.  These tori spectra are then convolved with the Doppler broadening appropriate for a disk in Keplerian motion (using the {\tt diskline} model with an emissivity profile $r^{-2}$) assuming an inclination of 45$^\circ$.  

\begin{figure*}[tbp]
\begin{center}
\includegraphics[width=4.3cm, angle=-90]{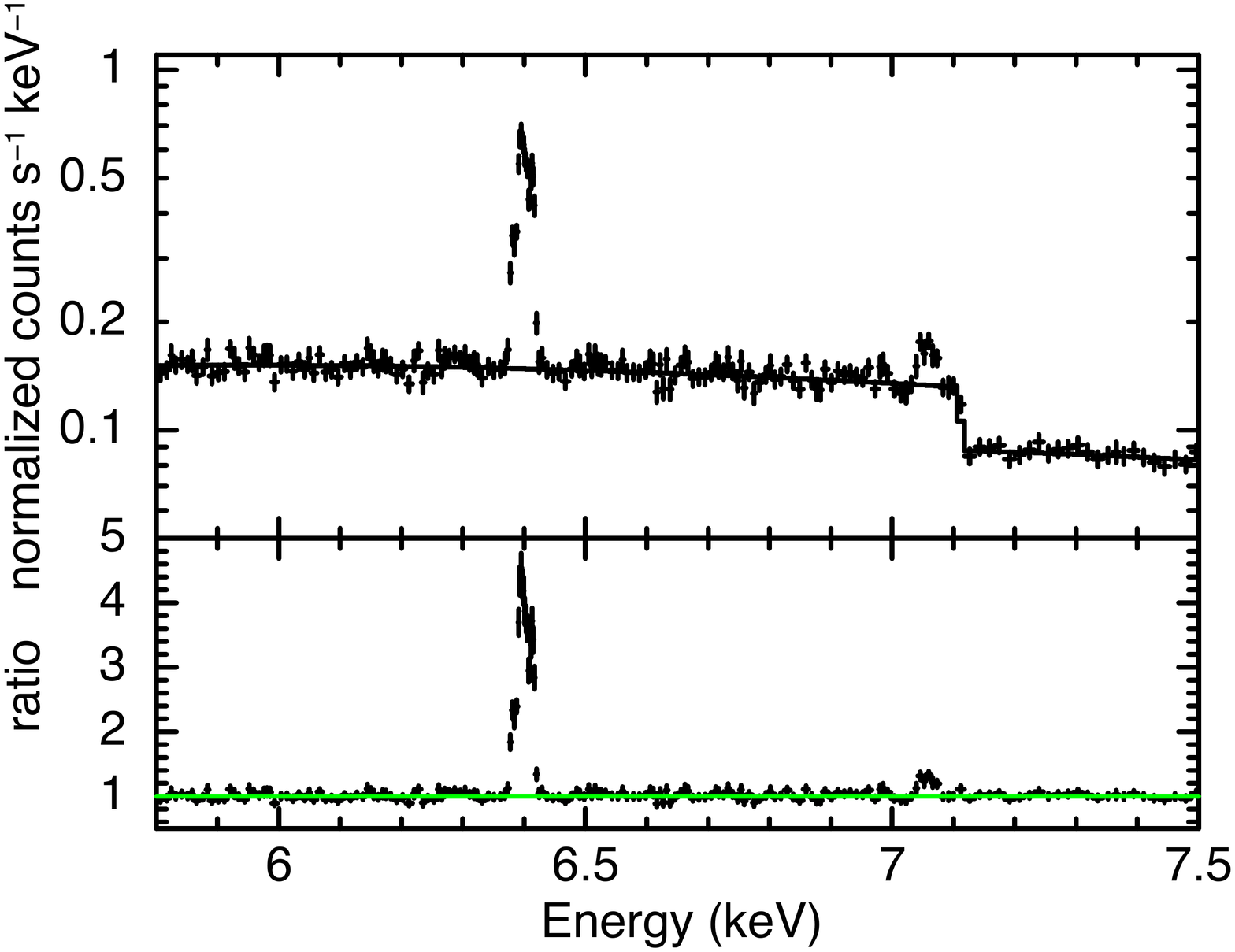}
\includegraphics[width=4.3cm, angle=-90]{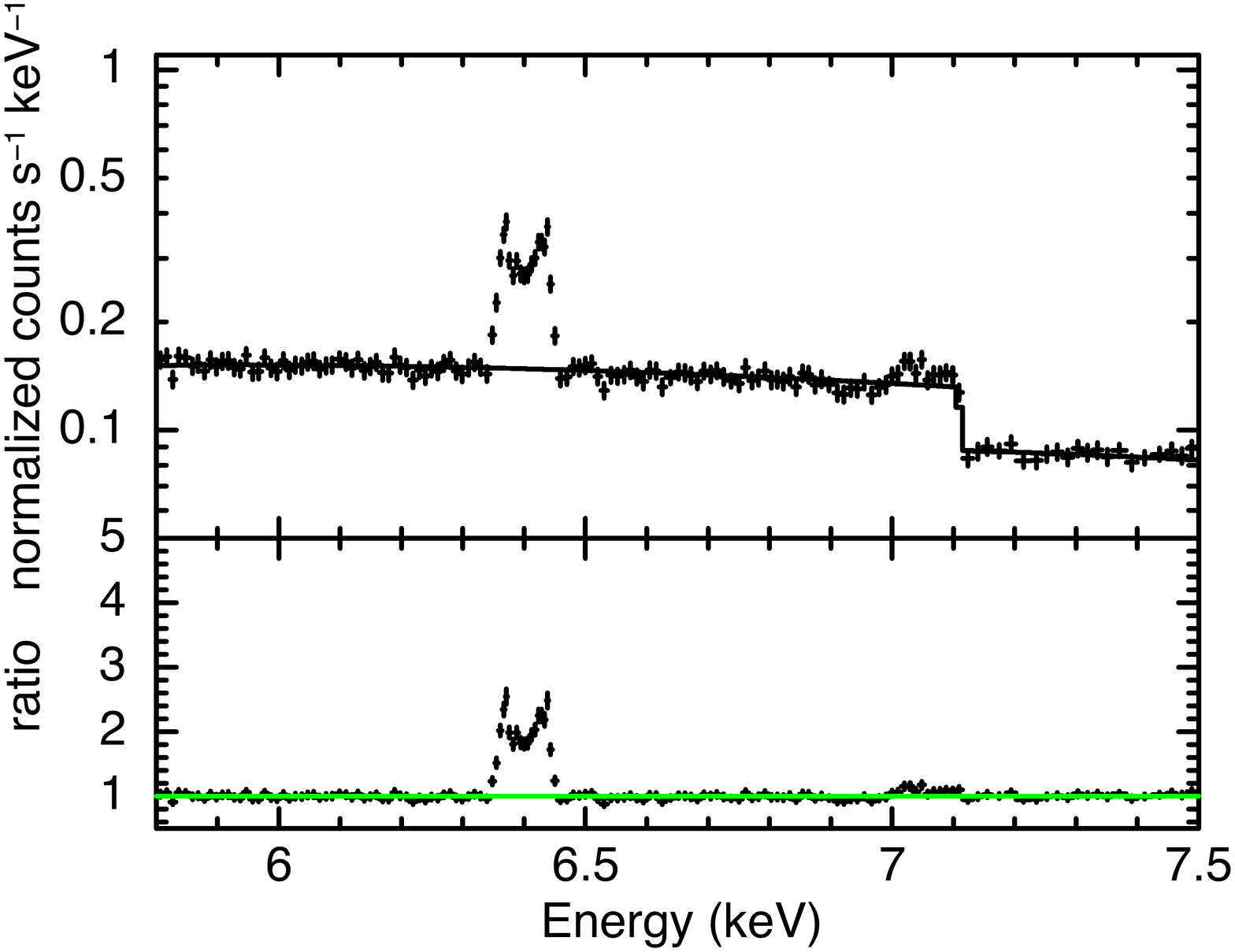}
\includegraphics[width=4.3cm, angle=-90]{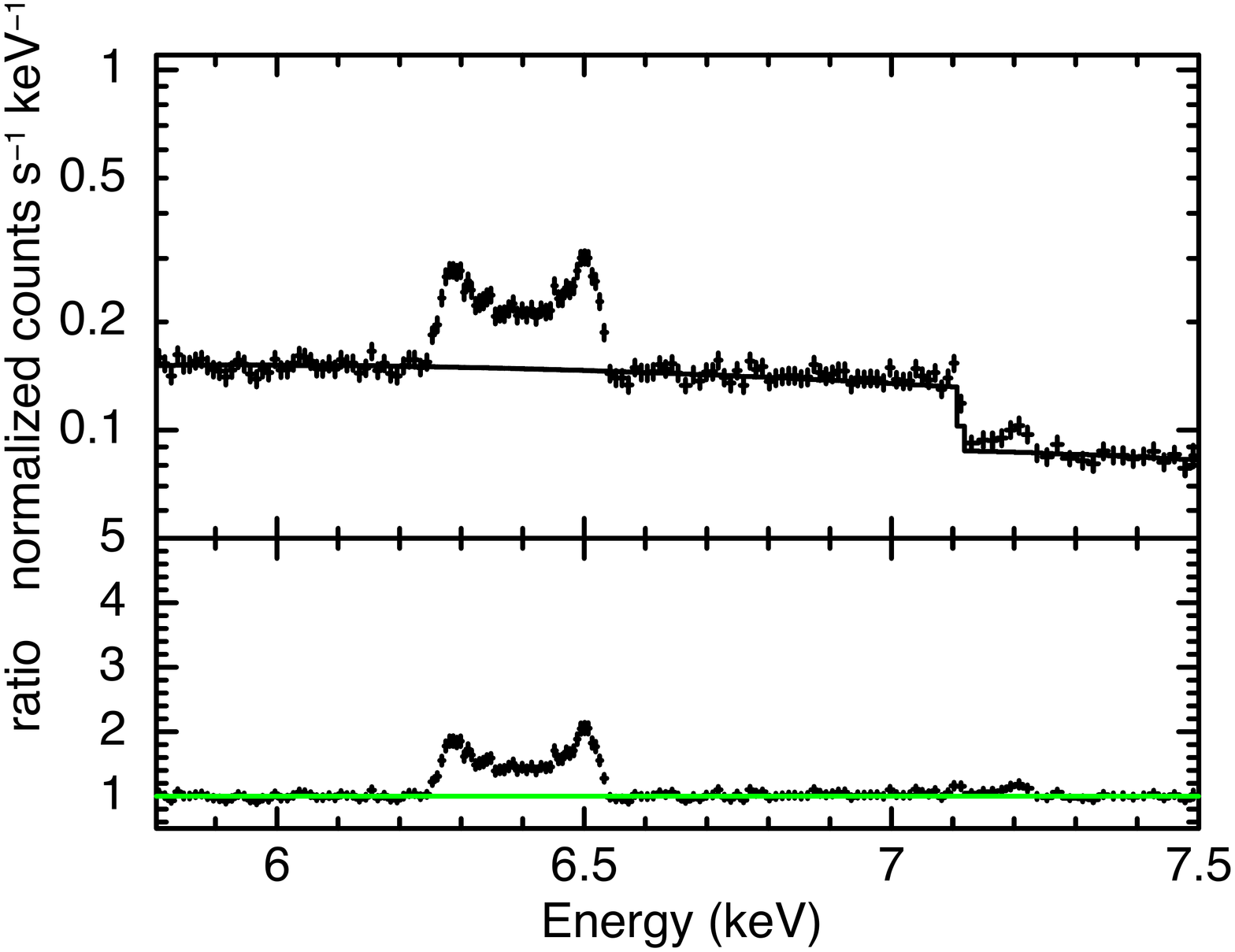}
\caption{Simulated spectra of NGC 4388 with the SXS, assuming 300\,ks exposure. Three cases shown are the far torus (left), near torus (middle) and BLR (right).}
\end{center}
\label{fig:ngc4388_felines}
\end{figure*}

Figure~4 shows a simulated 300\,ks SXS spectra in the iron-K band for NGC~4388, the brightest obscured Compton-thin AGN with a 14--195\,keV flux of $F=2.8\times 10^{-10}\rm\,erg\,s^{-1}\,cm^{-2}$ \citep{baumgartner12}. It is clearly seen that details of the 6.4\,keV line profile are resolved, in this case revealing the double-peak structure characteristic of the (assumed) keplerian disk. We can use the centroid and K$\alpha$/K$\beta$ line ratios to constrain the dominant ionization state of the iron; for the near torus case, we obtain the range Fe~II--Fe~VI, consistent with the XSTAR model that is dominated by Fe~II.  In this simulated case, we can also constrain the inner radius of the torus to better than 3\%, and measure mean bulk motion with a precision of $< 100$ km s$^{-1}$.  

While the parameters of this simulation are clearly simplistic, it highlights the fact that {\it ASTRO-H} will open a new window on AGN tori.  

\subsection{Broadband Continuum studies of Type-2 AGN with {\it ASTRO-H}}

The Cosmic X-ray Background (CXB) results from the integrated X-ray emissions of all the AGN in the Universe, and the spectrum, normalization and spatial fluctuations of the CXB are important constraints on models for the cosmological evolution of SMBHs.  While it is spectrally smooth, the CXB has a substantially harder spectrum than any known unabsorbed AGN and, in fact, absorbed AGN are crucial components any CXB model.  A good knowledge of the reflection and continuum parameters of AGN is fundamental for the creation of such models.  Especially important but ill-constrained are the cut-off energies for the power-law continua from AGN resulting from the finite temperatures of the X-ray emitting coronae.   This is another area where {\it ASTRO-H} will make a major impact, this time due to its ability to simultaneously determine the parameters of the primary continuum, scattering of that continuum, reflection, and the continuum cut-off.

Again, we illustrate this with the example of NGC~4388.  A broad-band study of this source
performed combining {\it XMM-Newton} EPIC/PN, {\it INTEGRAL} IBIS/ISGRI and {\it Swift}/BAT data shows an absorbed continuum with a photon index of $\Gamma=1.79^{+0.07}_{-0.08}$ and moderate levels of reflection (Ricci et al. in prep).   However, even with this array of data, the high-energy cutoff of the power-law could not be determined.  

To appreciate the importance of the high-energy cut-off, it is necessary to consider the physical processes responsible for producing the primary X-ray continuum.  The primary X-ray continuum is thought to be produced through Comptonization of the thermal optical/UV photons from the accretion disk by hot electrons in a magnetically energized corona (e.g.,\ \citealp{haardt93}).  The most basic parameters that we need to know about the corona if we wish to understand its nature and origin are its temperature $kT$ and optical depth $\tau$.  Now, the resulting spectrum is approximated as an exponentially cut-off power-law, with a cut-off energy at $E_C\sim 3kT$.  If we only have knowledge of the photon index of the power-law, we can only constrain the combination $kT\times \tau$ (proportional to the so-called Compton $y$-parameter).  Only if we measure the high-energy cutoff we can break the degeneracy between these fundamental coronal parameters.

\begin{figure*}
\centering
\begin{minipage}[!b]{.48\textwidth}
\centering
\includegraphics[width=8cm]{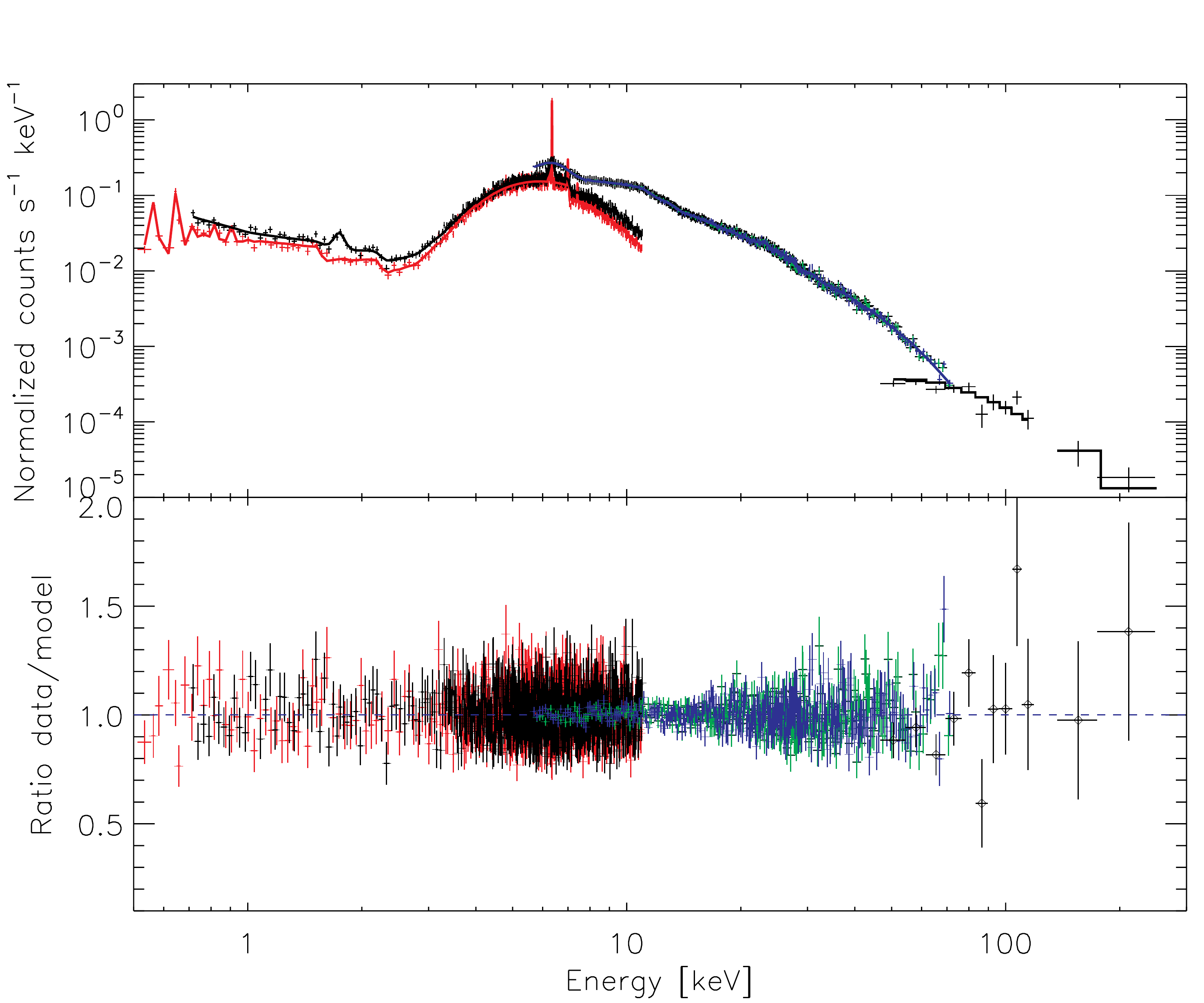}
\end{minipage}
\hspace{0.05cm}
\begin{minipage}[!b]{.48\textwidth}
\centering
\includegraphics[width=7cm,angle=90]{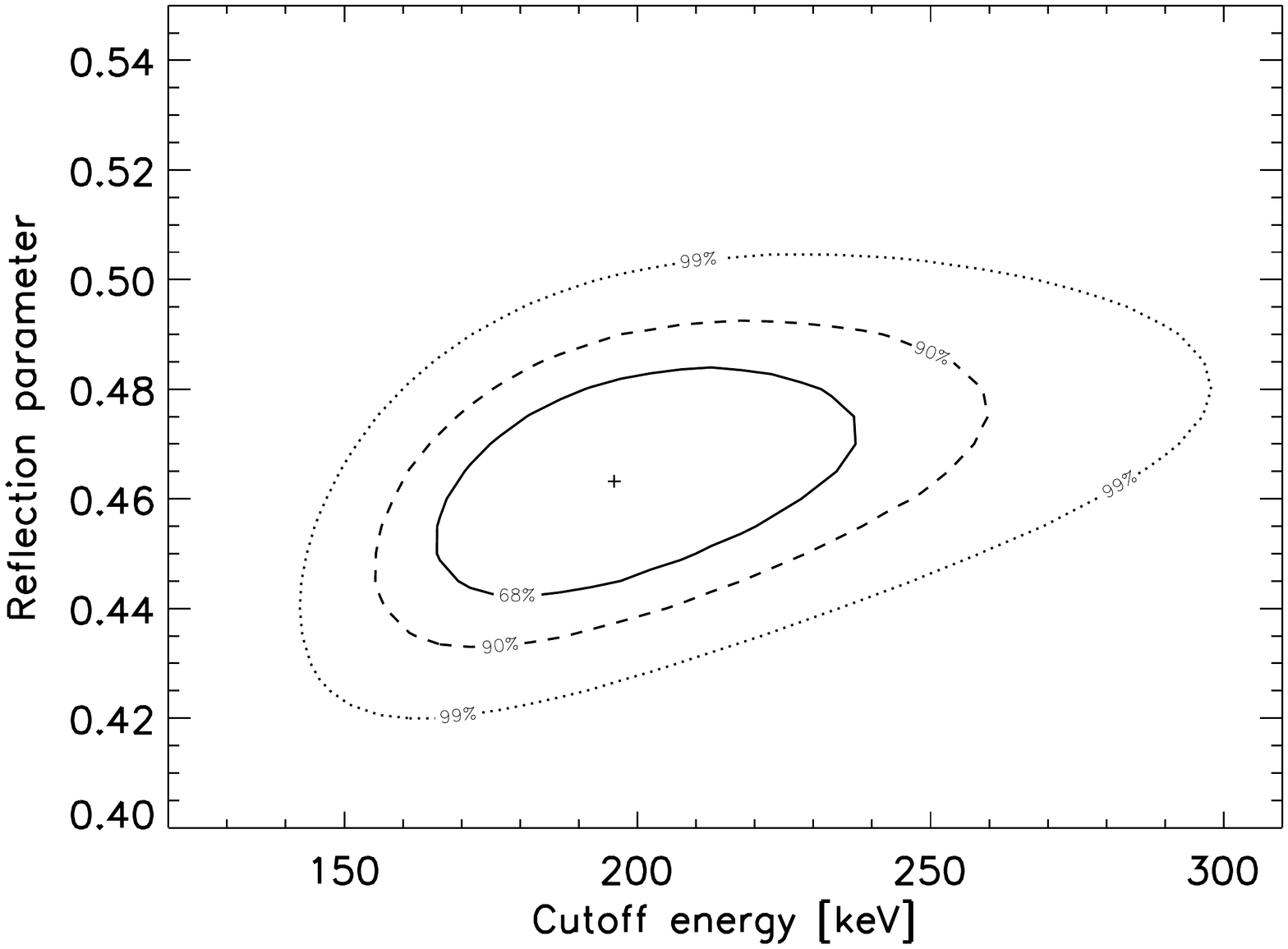}
\end{minipage}
 \begin{minipage}[t]{1\textwidth}
  \caption{{\it Left panel}: 100\,ks simulation of the broad-band {\it ASTRO-H} spectrum of NGC\,4388.  Shown here are data from the SXS (red), SXI (black), HXI (blue and green), and SGD (magenta).  {\it Right panel}: contour plot of the reflection component versus the energy of the cutoff obtained by fitting the simulated 100\,ks {\it ASTRO-H} observation of NGC\,4388. The cross represents the best set of parameters obtained by the fit, while the three contours are the 68\%, 90\% and the 99\% confidence regions.}
\label{fig:NGC4388_broad-band}
\end{minipage}
\end{figure*}

Very recent {\it NuSTAR} results are already demonstrating the dramatic impact that a focusing hard X-ray telescope is having on our abilities to measure cut-off energies in AGN spectra (see study of IC4329A; Brenneman et al. in prep).  {\it ASTRO-H} will take these studies to the next level, providing {\it NuSTAR} quality spectra from the HXI with simultaneous soft X-ray data from the SXI/SXS and harder data from the SGD.  Figure~\ref{fig:NGC4388_broad-band} (left) illustrates this with a 100\,ks multi-instrument simulation of an {\it ASTRO-H} observation of NGC~4388.  In a single observation, the spectral shape can be determined with high precision from 0.5--80\,keV, and the SGD detection gives us an important lever-arm out to $\sim 200$\,keV on our ability to determine cut-off energies.  In this simulation, we assumed a reflection strength (measured relative to that obtained from a source above an infinite slab) of $R=0.46$, and cut-off energy of $E_C=200$\,keV.  As can be seen in Figure~\ref{fig:NGC4388_broad-band} (right), the {\it ASTRO-H} multi-instrument dataset allows us to break any degeneracy between $R$ and $E_C$, and measure each to a high precision; $R=0.46\pm 0.02$ and $E_C=196^{+42}_{-31}$\,keV.   The constraints on the cut-off energy are systematically improved as that energy is lowered and more of the curvature is placed into the HXI band, with 12\% errors if $E_C=100$\,keV and 7\% errors for $E_C=50$\,keV.   However, even for $E_C=500$\,keV, we can still detect the cut-off and measure its energy to $\pm 40$\%.

\begin{figure*}[t]
\centering
\includegraphics[width=8cm]{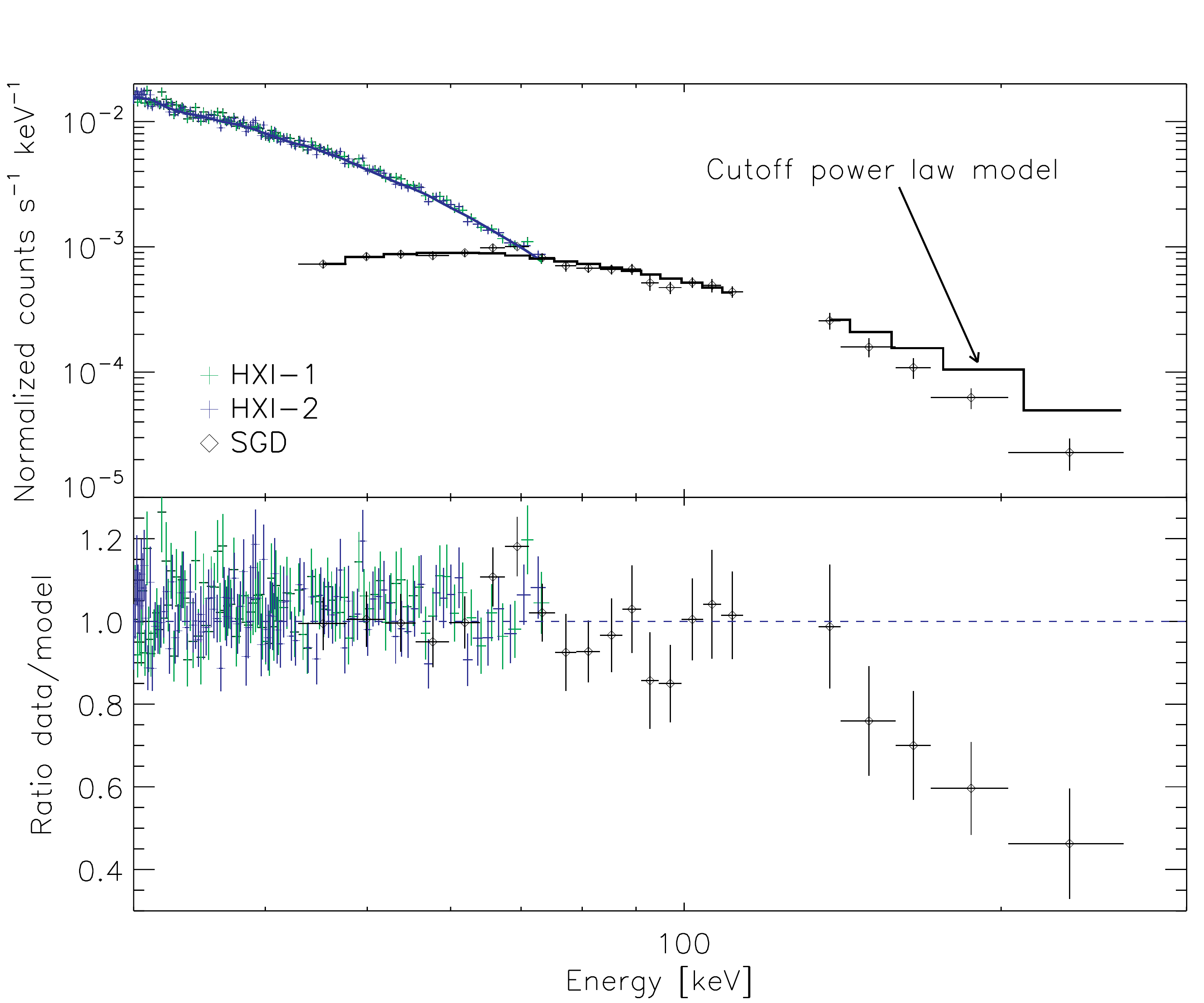}
 \caption{Extract of the spectrum obtained by a 100ks {\it ASTRO-H} observation of NGC\,4388 simulated using a Comptonization model for the continuum ($T_{\mathrm{e}}=50$\,keV, $y=1$). The simulated spectrum was fitted using a cutoff power-law to account for the primary X-ray emission, which, as it can be seen from the bottom panel, overestimates the flux above 150\,keV.}
\label{fig:NGC4388_compps_cutoff}
\end{figure*}

An exponentially cutoff power-law is only a first approximation to the X-ray continuum of AGN.    The precise form of the spectrum produced by a physical thermal Comptonization model  can be crudely approximated by an exponential power-law across a certain range of energies, but will deviate significantly from that simple model for $E>E_C$.  Once we are sensitive to these deviations, we have opened a new window on the geometry and structure of the corona.  Here, the combination of the HXI and SGD on {\it ASTRO-H} will make a qualitative advance.   We illustrate this for the case of NGC~4388 in Figure~\ref{fig:NGC4388_compps_cutoff}; we simulate a 100\,ks multi-instrument spectrum using a physical Comptonization model (\texttt{compPS}, \citealp{poutanen96}) but fit the simulated data with an exponential powerlaw.  Assuming a coronal temperature of $50$\,keV, we can see that dramatic and easily measurable deviations from the exponential cutoff power-law model are seen above 150\,keV.

\section{The Nature of Compton-Thick AGNs}\label{sec:comptonthick}

\subsection{The Importance of Compton-Thick AGN}

The shape of the CXB requires that a significant population of highly obscured AGN must exist.  At the extreme end of the distribution lies the Compton-Thick AGN (CTAGN), those for which the electron scattering optical depth to the central engine exceeds unity implying gas column densities exceeding $N_{\rm H}\sim 10^{24}$ cm$^{-2}$.   For obvious reasons, this makes them difficult to find --- we must either look in the highest energy X-ray bands (some fraction of which can still penetrate the obscurer unless $N_H>10^{25}\,{\rm cm}^{-2}$) where sensitivity has historically been challenging or search for secondary/reprocessed signatures.  Their unknown prevalence represents the most important remaining factor affecting our understanding of the composition of the CXB \citep{ueda03,gilli07}, and is a major uncertainty in our understanding of the coevolution of galaxies with their SMBHs.  To put it bluntly, our knowledge of up to half of the accretion energy budget of the universe remains hidden from view, with our current knowledge being extrapolations from biased data.

Being extreme objects, CTAGN are also interesting in their own right.  It is unknown whether they are a disjoint population, possibly corresponding to some given phase in the life of a merger-driven AGN, or simply one end of the distribution of normal AGN.   Neither do we know their cosmological evolution.  Hierarchical growth models posit that the nuclei of all large galaxies develop under large (probably Compton-Thick) columns of gas that feed as well as obscure the growing SMBH \citep{fabian99,hopkins06}. This implies that the CTAGN abundance ought to evolve positively with redshift, and recent results appear to support this \citep[e.g.,][]{brightman12}. Changing preliminary evidence into proof requires secure identification of statistically-significant samples of CT AGN, as well as detailed spectral follow-up to measure physical parameters such as their covering factor.

Here, we discuss the impact of {\it ASTRO-H} on our understanding of the physical nature of CTAGN.  Again, this comes both from the high-spectral resolution of the SXS and the remarkable broad-band capabilities of {\it ASTRO-H}.

\subsection{The Compton Shoulder as a Probe of the Obscurer}

\begin{figure*}[htdp]
\centering
\includegraphics[width=0.6\hsize]{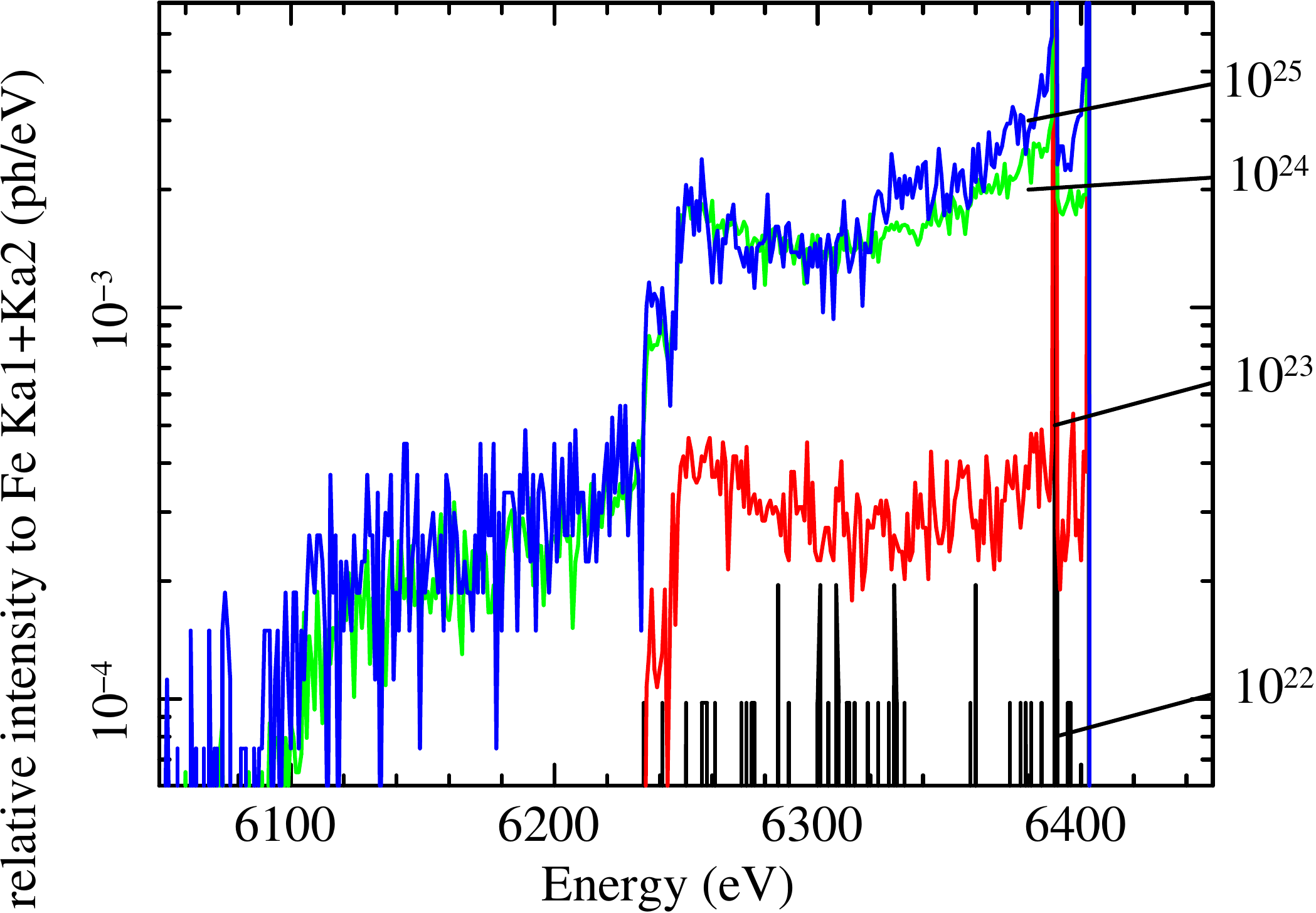}
\caption{Monte Carlo simulations of Compton shoulders for various column densities.  A torus geometry is assumed following \citet{ikeda09}, with a half opening angle of $30^\circ$ and a viewing angle of $30-32^\circ$ (i.e., grazing the edge of the torus).  For simplicity, the scatters are assumed to be cold free electrons.}
\label{fig:rel_int}
\end{figure*}

As explained in Section~1, iron fluorescence results when cold matter is irradiated by a hard X-ray continuum.   If the irradiated structure is Compton-Thick, iron line photons that are produced heading {\it into} the interior can be Compton scattered back towards the surface and, with some probability, to the observer.  Due to Compton-recoil, these scattered iron line photons are downshifted in energy by $\Delta E=0-0.15\keV$ and form a ``Compton shoulder'' sitting redwards of the main line.   In fact, there is a series of Compton shoulders corresponding to iron line photons that have been once-scattered, twice-scattered etc., each becoming successively downshifted and broadened compared with the previous (see Figure~\ref{fig:rel_int}).  

The detailed strength and shape of the main shoulder is a novel probe of the geometry and conditions of the reflector.   Its detailed shape can be used to determine if the scattering electrons are bound in atoms/molecules or free (see also White Paper 17), and whether the observed iron line does in fact come from a Compton-Thick structure, even if the line-of-sight absorption is Compton-thin.  Calculations of the type shown in Figure~\ref{fig:rel_int} reveal that the relative strength of the Compton shoulder decreases with increasing the half opening angle of the Compton thick torus.  Thus we can tell whether an AGN is deeply buried, i.e. almost surrounded by Compton-Thick material, by looking for a strong Compton shoulder.  

\begin{figure*}[htdp]
\centering
\includegraphics[width=0.7\hsize]{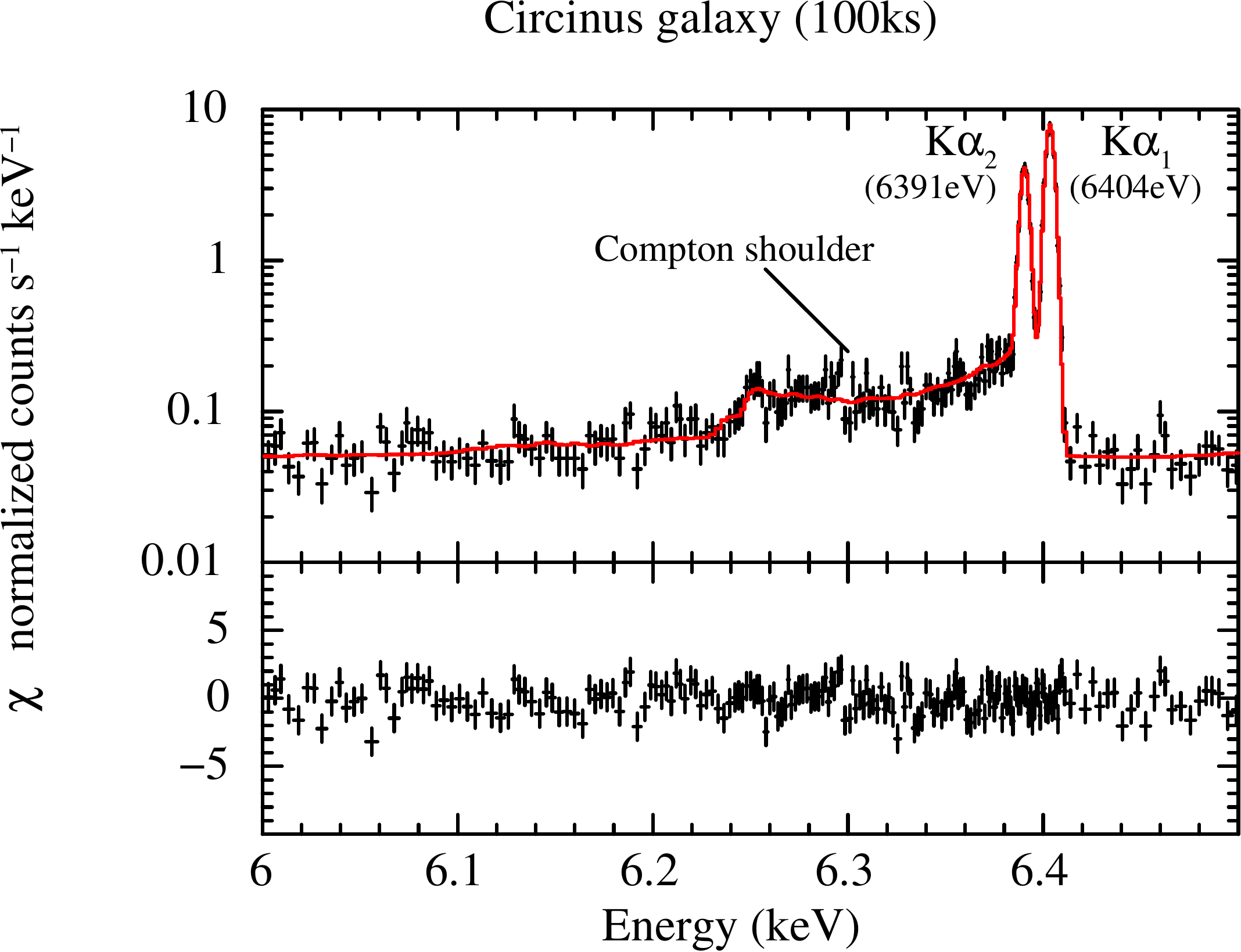}
\caption{Simulated SXS spectrum of the Compton-Thick Circinus galaxy in the
iron-K$\alpha$ band. The exposure is 100~ks.}
\label{fig:cir_100k}
\end{figure*}

While hints of the Compton shoulder have been seen in {\it Chandra} grating data, the {\it ASTRO-H}/SXS is the first instrument capable of truly utilizing this diagnostic.  Figure~\ref{fig:cir_100k} shows a simulated SXS spectrum from a 100\,ks observation of the Circinus galaxy, assuming a spectral model with a Compton-Thick torus, and a half opening angle of 30$^\circ$.  Thanks to the intense iron line in this confirmed CTAGN, the Compton shoulder will be clearly observed.

\subsubsection{Broadband Continuum Studies of CTAGN}

\begin{figure*}[htdp]
\centering
\begin{minipage}[!b]{.48\textwidth}
\centering
\includegraphics[width=1.0\hsize]{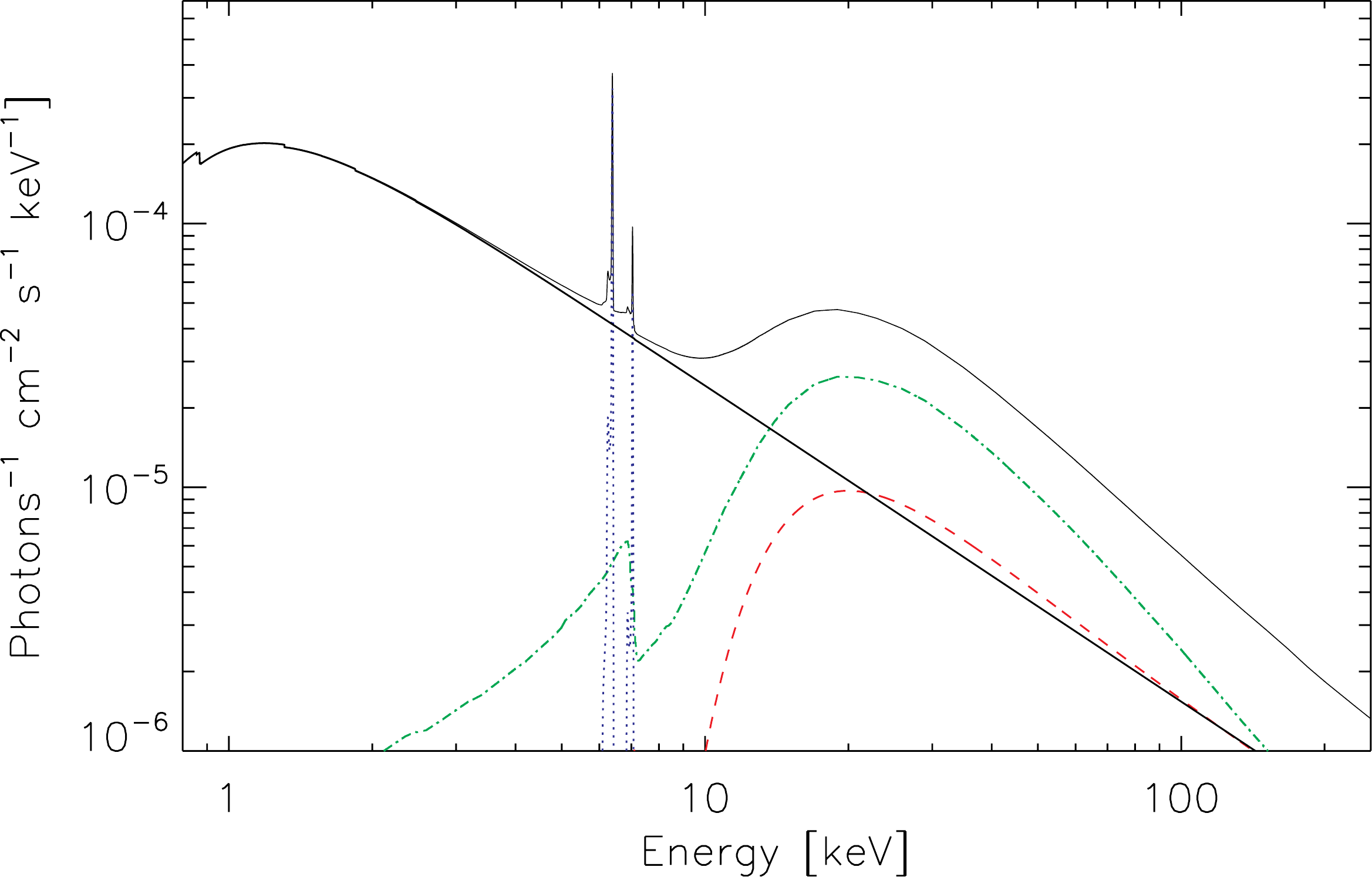}
\end{minipage}
\hspace{0.05cm}
\begin{minipage}[!b]{.48\textwidth}
\centering
\includegraphics[width=1.0\hsize]{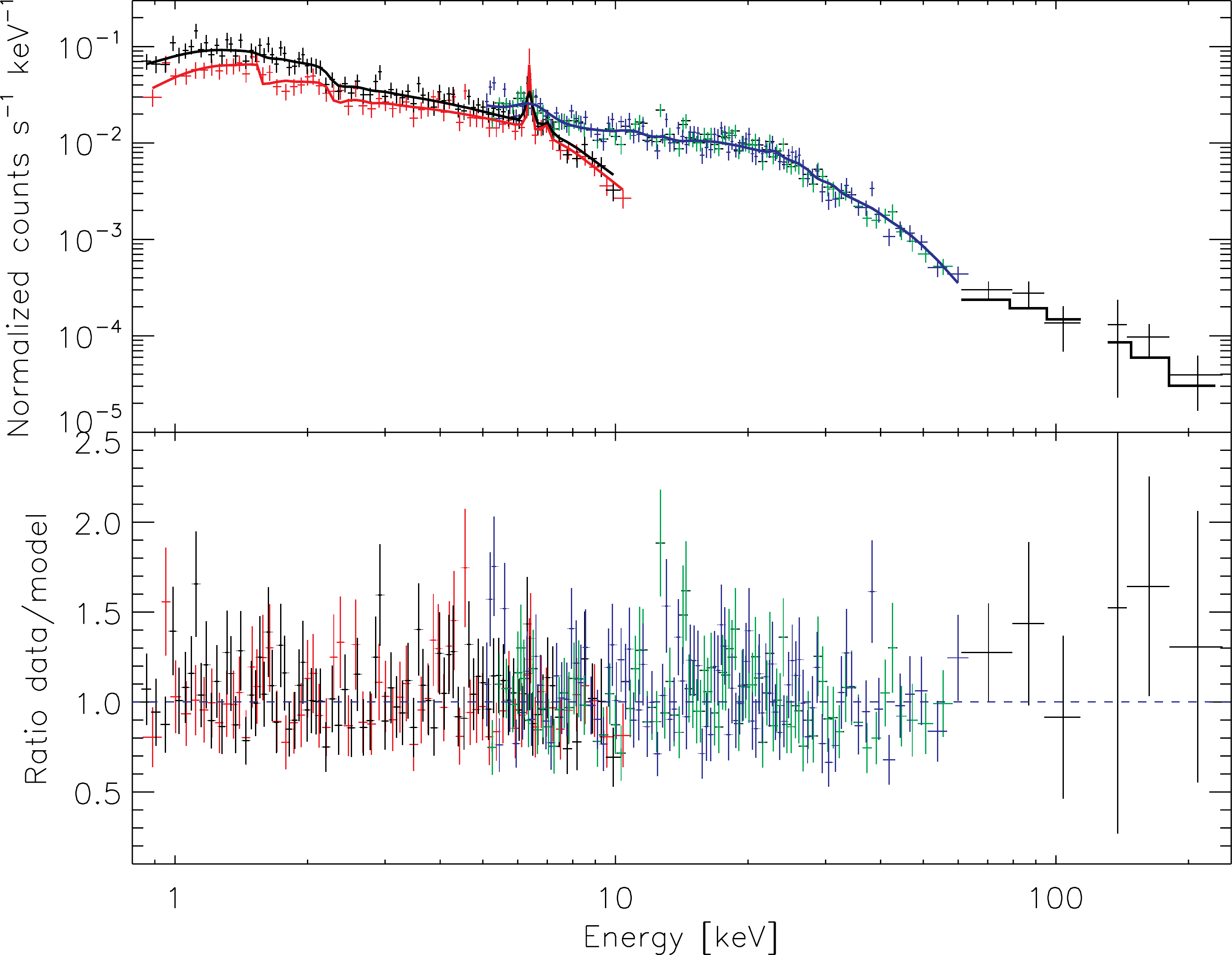}
\end{minipage}
\caption{{\it Left:} Spectral model for NGC~4945. {\it Right:} Simulated SXS (red), SXI (black) and HXI (green) spectra of NGC 4945, with a 10\,ks exposure.
}
\label{fig:NGC4945_spec}
\end{figure*}

For understanding the nature of a CTAGN, it is critical to accurately constrain the column density and its intrinsic luminosity. With its unique combination of broad band-pass and sensitivity, {\it ASTRO-H} will permit unprecedented studies of the continuum in CTAGN.  We illustrate this with the case of NGC~4945, a bright AGN with a line-of-sight column of $N_H=4\times 10^{24}\,{\rm cm}^{-2}$ which completely extinguishes any primary continuum below 10\,keV.  Prior studies have shown that the X-ray spectrum is dominated by scattered continuum (below 10\,keV) and a reflection component (above 10\,keV), with the absorbed primary continuum remaining subdominant out beyond 50\,keV (see Figure~\ref{fig:NGC4945_spec}, left).   

Figure~\ref{fig:NGC4945_spec} (right) shows simulated SXS/SXI/HXI spectra from a 10\,ks {\it ASTRO-H} observation of NGC~4945. It is clear that the continuum shape can be defined even in such a short observation; the photon index can be measured to $\pm 0.2$ and the column density can be constrained with 20\% errors.   While NGC~4945 is very local ($z=0.0019$) and extremely bright, similar constraints can be obtained for a typical CTAGN at $z=0.01$ with a 200\,ks exposure.

\section{Probes of the Relativistic Accretion Disk and SMBH}
\label{sec:relrefl}

\subsection{Into the Heart of the AGN}
\label{sec:relref_background}

\begin{figure*}
\centerline{
\includegraphics[width=0.5\hsize]{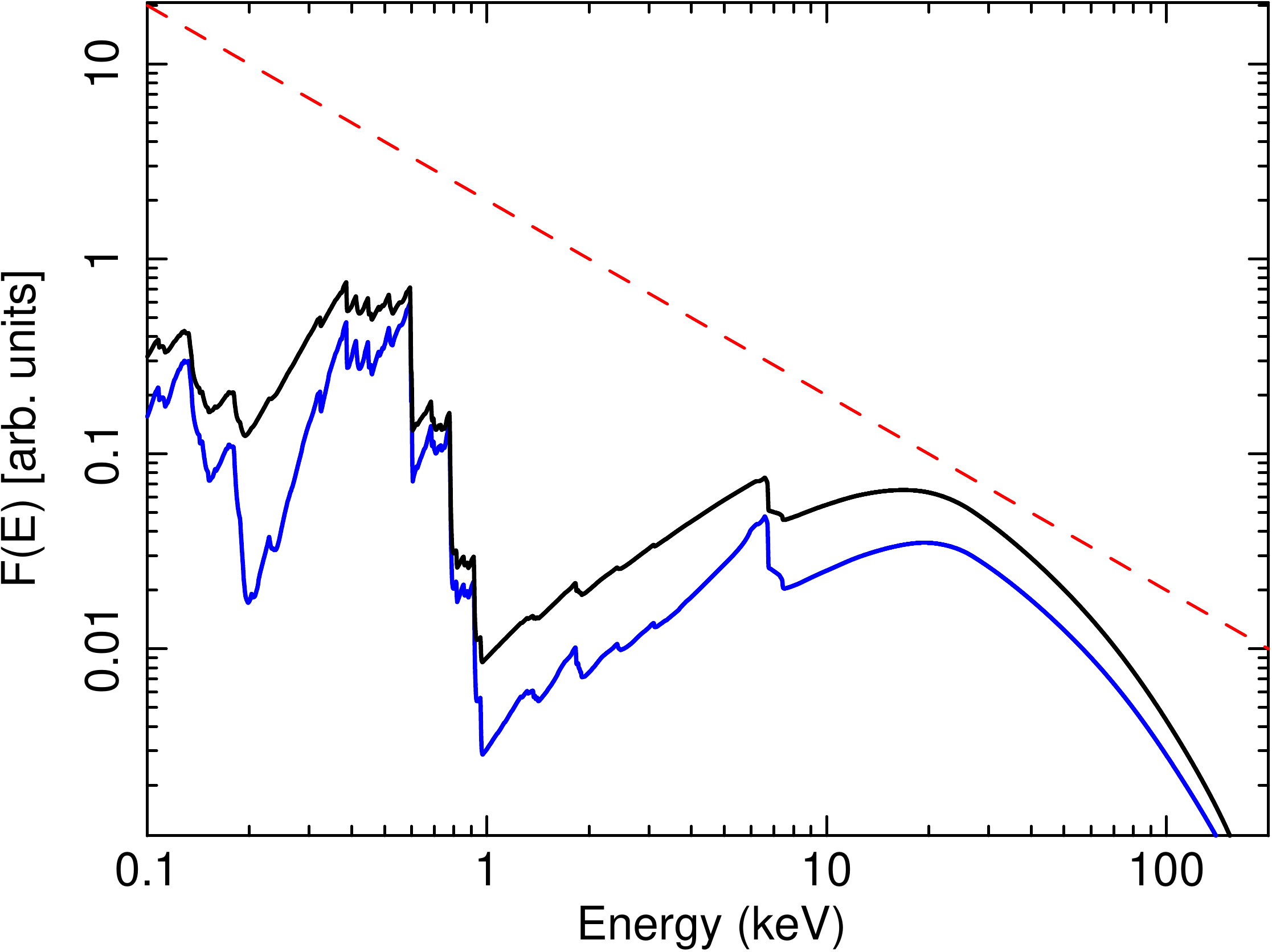}
}
\caption{Effects of relativistic Doppler and gravitational redshift effects on an ionized reflection spectrum assuming a viewing inclination of $i=30^\circ$, irradiation index of $q=3$, and a rapidly spinning BH ($a=0.99$; black line) or non-spinning ($a=0$; blue line) BH.   Figure from \cite{reynolds13}.}
\label{fig:blurredrefl}
\end{figure*}

In order to understand the AGN phenomenon, we must ultimately understand the physics within a few gravitational radii of the black hole where the vast majority of the energy is released.   X-ray irradiation of the ionized inner regions of the accretion disk, and the associated X-ray reflection, gives us our cleanest probe to date of this region.  As illustrated in Figure~\ref{fig:blurredrefl} Spectral features in the reflected X-ray emission are strongly broadened and skewed by the Doppler effect from the accretion flow's orbital motion as well as gravitational redshifts as the photons climb out of the deep potential well \citep{tanaka95,fabian00,reynolds03}.  In addition to producing the well-known broad iron line, ionized reflection from the inner disk can also produce at least some part of the soft excess as well as a broadened Compton reflection hump.

To study the inner accretion disk, we must disentangle these spectral features from other complexity.   Even in sources where the line-of-sight to the inner disk is Compton-thin, there may still be significant absorption features imprinted into the spectrum from partially photoionized material (the so-called warm absorber).   As discussed in the first part of this WP, emission signatures from scattering/reflection of X-rays by distant (low velocity) matter can further contribute to and complicate the spectra.   As we will describe below, {\it ASTRO-H} will provide powerful new ways of disentangling these complexities from the accretion disk features.

The primary X-ray source is highly variable.  So, in addition to the spectrum, timing provides another way to probe the heart of the AGN.  Particularly exciting is the discovery of reverberation time delays in the {\it XMM-Newton} and {\it Suzaku} data of some bright AGN.   It has long been predicted that, since the X-ray source is external to the accretion disk, there will be light travel induced time-lags between the observed continuum changes and the observed response of the reflection \citep{reynolds99}.  Using {\it XMM-Newton} and {\it Suzaku}, these reverberation delays have now been seen in the (soft) reflection spectrum of several narrow line Seyfert 1 galaxies \citep{zoghbi10,demarco13}, as well as in the broad iron line of several hard X-ray bright AGN \citep{zoghbi12}.

\subsection{Disentangling the Accretion Disk Spectrum From Other Complexities}

The disk reflection spectrum is a broad-band component, producing a soft excess, a broad iron line and a hard X-ray bump.   When extracting accretion disk (or SMBH spin) parameters from an X-ray spectrum, it is crucial to know the X-ray continuum that is superposed on this disk reflection spectrum as well as the effects of any absorption.    {\it ASTRO-H} is a powerful observatory for disentangling even complex AGN spectra.   The SXI and HXI provide a high signal-to-noise and broad-band view of the spectrum, allowing us to simultaneously study the continuum, soft excess and Compton hump.   For bright objects or long exposures, the SXS provides a high-fidelity view of narrow emission and absorption lines, removing any influence that such features may have on the modeling of the underlying broadened disk features.  Taken all together, {\it ASTRO-H} will give us our best ability to date to study broad iron lines and relativistic disk reflection in a manner that is robust to the other spectral complexities.  

\begin{figure}[t]
\centerline{
\includegraphics[width=0.6\hsize]{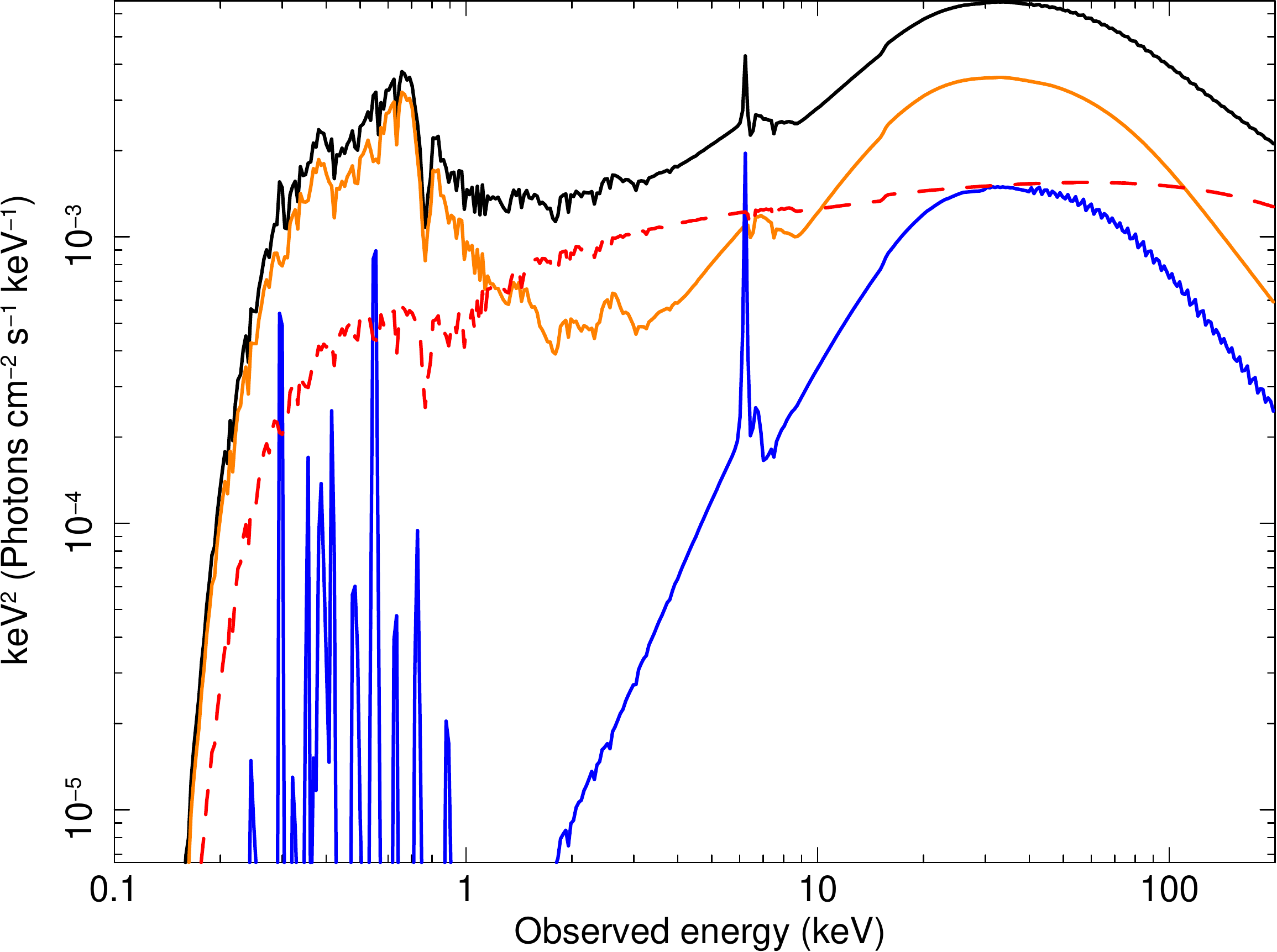}
}
\caption{The spectral model adopted for the simulations is similar to that of Mrk~335 in the intermediate flux state.  The model is composed of a power law continuum, a blurred reflection component (blue line), and a distant reflector.  The broad-band continuum is modified by a three-zone warm absorber and a Galactic column density.}
\label{fig:mrk335_mod}
\end{figure}

\begin{figure}[t]
\centerline{
\includegraphics[width=0.7\hsize]{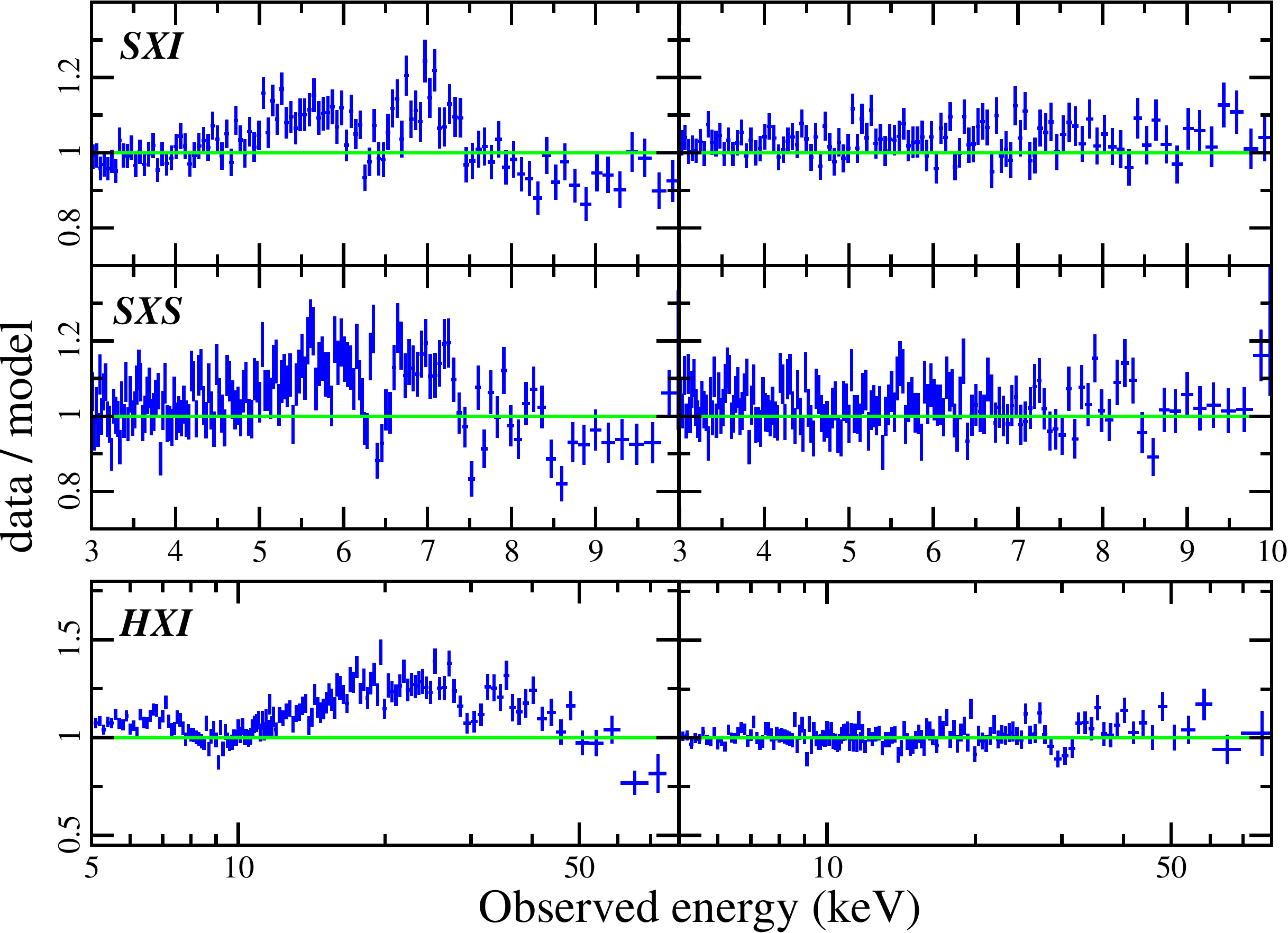}
}
\caption{A power law plus distant reflector model is fitted to the 200 ks simulated spectra of an AGN with $F_{2-10keV} = 2.5\times 10^{-11} \ergpcmsqps$.  The residuals in each instrument are shown on the left.  The broad component becomes visible in all detectors and the SXS begins to reveal the structure of the warm absorber.  A blurred reflector and warm absorber are then added to the model.  The residuals from the final fit are shown on the right.}
\label{fig:mrk335_ratio}
\end{figure}

We illustrate the ability of {\it ASTRO-H} to disentangle warm absorption, torus reflection and disk reflection by adopting the spectrum of Mrk~335 ($z=0.026$) as a representative of  the ``typical'' type-1 Seyfert population.  Using {\it XMM-Newton} data, \cite{gallo13} find that this object displays a complex warm absorber (i.e. multiple zones of photoionized absorption) and reflection from a torus, in addition to the relativistically blurred reflection from the central regions of a moderately ionized, $\xi=200\erg\cmps$, accretion disk (see Figure~\ref{fig:mrk335_mod}).   We simulate a 200\.ks {\it ASTRO-H} observation of this spectrum, assuming a 2--10\,keV source flux of $2.5\times 10^{-11}\ergpcmsqps$; this is brighter than is typical for Mrk~335, but well within the flux range for many other similar type-1 Seyfert galaxies.  

Figure~\ref{fig:mrk335_ratio} (left panels) shows the residuals that remain when we fit the $>3$\,keV spectrum with a model consisting of the power-law and torus reflection but does not include the warm absorber or the accretion disk.  The unmodelled broad iron line can be clearly seen in both the SXI and SXS residuals, and the unmodelled Compton reflection from the accretion disk is apparent in the HXI residuals.  Furthermore, sharp narrow absorption lines in the SXS result from the unmodelled warm absorption.  When the full spectral model is used, the residuals are flat as expected (Figure~\ref{fig:mrk335_ratio}, right) and we can obtain excellent constraints on the parameters of the relativistic disk --- we recover the inner edge of the disk to $\Delta r_{\rm in}= 0.3r_g$ (input $r_{\rm in}=3r_g$), the inclination to $\Delta i=3^\circ$ (input $i=49^\circ$), the irradiation index to $\Delta q=0.4$ (input $q=5$) and the disk iron abundance to $\Delta Z=0.2Z_\odot$ (input $Z=1.9Z_\odot$).  The disk parameters can be recovered despite the presence of the highly ionized absorber that affects the iron band.  

An interesting point to note is the relative roles played by the SXI, SXS and HXI.    While contributing significant signal-to-noise in the detection of broad features in the 3--10\,keV band, the principal role of the SXS is to detect and characterize absorption features, thereby removing any remaining degeneracies in the fits to the disk reflection arising from uncertainties in absorption.    Using just the soft X-ray detectors (the SXI and SXS), we can detect the presence of blurred reflection but cannot statistically distinguish the actual blurring parameters ($R_{\rm in}=3\rg, q_{\rm in}=5, q_{\rm out}=3, R_{\rm br}=6\rg, i=49^\circ$) from some default set ($R_{\rm in}=4.5\rg, q_{\rm in}=q_{\rm out}=3, i=30^\circ$).  It is only with the inclusion of the HXI data that the blurring parameters can be constrained in detail.  This follows from the modest effective areas of the SXI and SXS compared with the HXI.  

The uncertain nature of the soft excess can also be a source of uncertainty when attempting to constrain the accretion disk parameters.  For example, \cite{lohfink12} examined multi-epoch {\it XMM-Newton} and {\it Suzaku} data for the unabsorbed Seyfert nucleus in Fairall~9 and found that, while the presence of relativistically blurred disk reflection was robust, the inferred black hole spin and disk iron abundance depended strongly upon whether the soft excess was model as an additional ionized blurred reflection (beyond that needed to probe the iron line) or thermal Comptonization from lukewarm material.   Part of the issue is that the broad line in Fairall~9 is not particularly strong compared to sources such as MCG--6-30-15 or NGC3783 --- however, it may be representative of ``normal'' sources with regards to its broad line strength.   

\begin{figure}[t]
\centerline{
\includegraphics[width=0.7\textwidth]{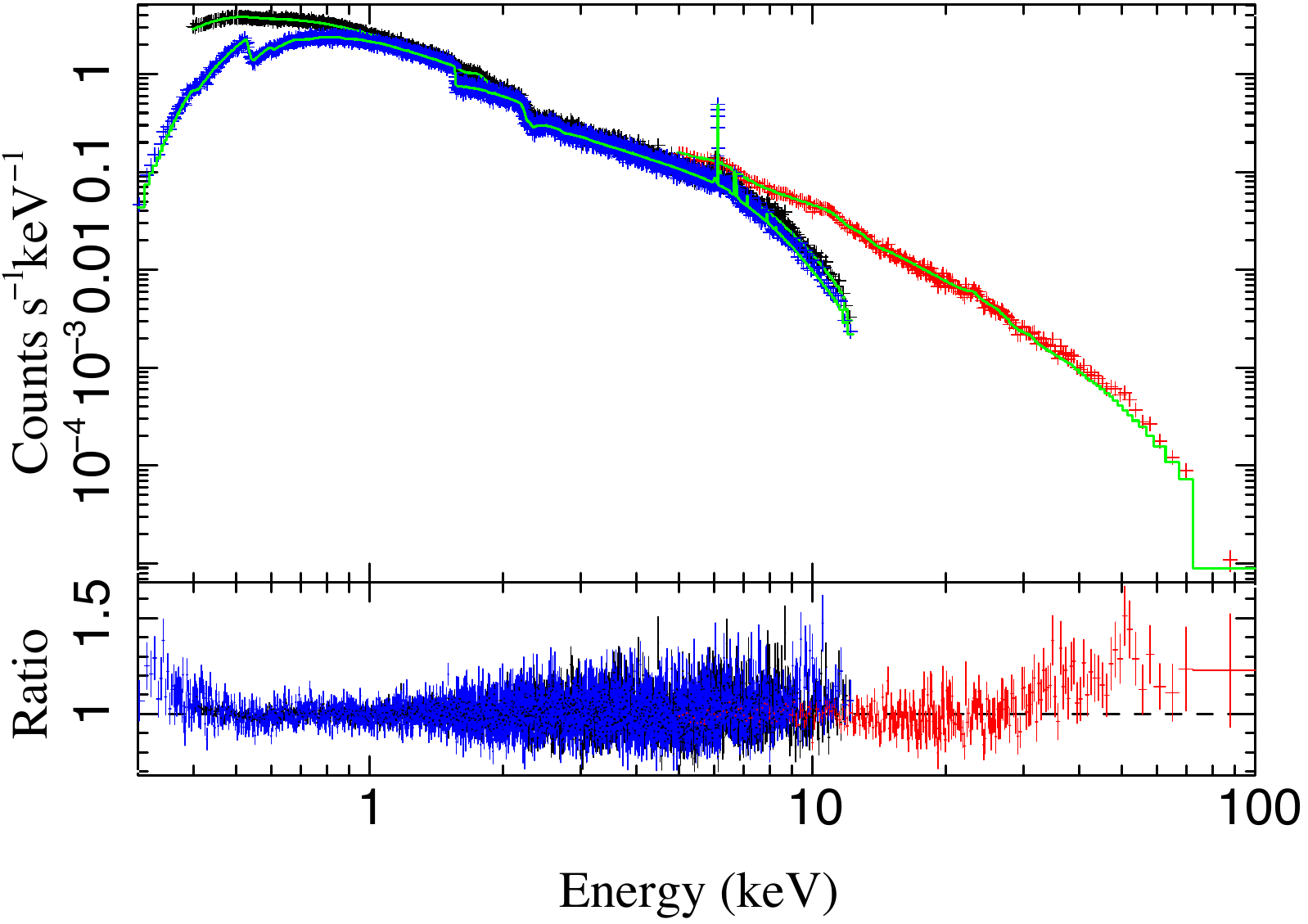}
}
\caption{Simulated 150 ks spectra of the Seyfert I galaxy Fairall 9, assuming a model with cold and ionized reflection, as well as a Comptonization soft excess. Shown here is the fit to the SXS [blue], SXI [black], HXI [red] spectra with a double ionized-disk-reflection scenario but no luke warm Comptonization component.  The precise form of the residuals clearly shows that the two scenarios can be distinguished.}
\label{fig:fairall9}
\end{figure}

{\it ASTRO-H} will be able to slice through the degeneracies responsible for the current ambiguities in a Fairall~9 like source.  To explicitly demonstrate this, we have adopted the ``Comptonization soft excess'' scenario for Fairall~9 and simulated 150\,ks exposures with the SXI, SXS and HXI, assuming a typical 0.5--10\,keV flux level of $4\times 10^{-11}\ergpcmsqps$.  To maximize realism, we also redshift the spectrum to $z=0.047$, the redshift of Fairall~9 (and quite typical for many Seyfert galaxies that we will wish to study).   Figure~\ref{fig:fairall9} shows a fit to these simulated spectra with the ``wrong'' soft-excess model, namely an additional blurred ionized disk component on top of that needed to produce the iron line. Residuals clearly remain at the softest and hardest energies, showing that {\it ASTRO-H} can resolve this tough degeneracy.

\subsection{X-ray Reverberation from the Inner Disk}

\begin{figure}[t]
\centerline{
\includegraphics[width=0.95\textwidth]{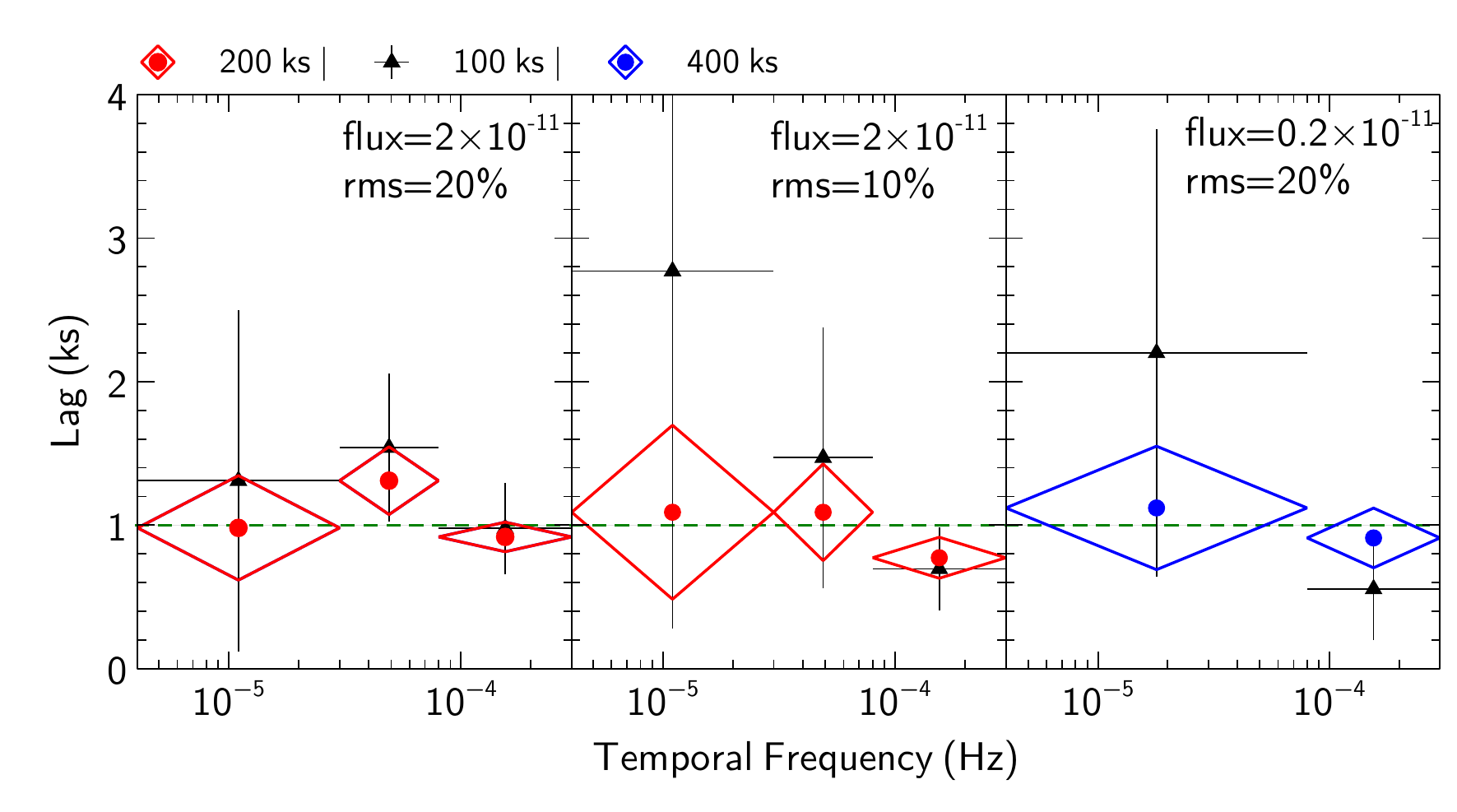}
}
\caption{Recovered time-lags in simulated HXI data.   The input simulation assumes a 1\,ks time lag (constant for all Fourier frequencies) between the continuum band (7--10\,keV) and the hard band containing the time-delayed Compton hump (20--50\,keV).  The simulation includes gaps in the lightcurve appropriate for a low-Earth satellite, and the quoted exposure time is the ``on source'' time.  The time lags are recovered using Fourier techniques that incorporate a maximum likelihood technique to handle the data gaps (Zoghbi et al. in prep).}
\label{fig:lags}
\end{figure}

As discussed in \S\ref{sec:relref_background}, {\it XMM-Newton} and {\it Suzaku} studies of bright AGN have recently opened an exciting new frontier in the study of reverberation effects from the inner accretion disk.  Here, we assess the impact that {\it ASTRO-H} will have on these studies.  

The total number of detected photons, and hence the effective area of the {\it ASTRO-H} instruments, is key to whether reverberation lags can be detected.  To formally assess detectability, we have simulated pairs of light curves assuming a standard spectral shape (Figure~\ref{fig:mrk335_mod}) with a fixed time-delay between the direct power-law and the reflected disk emission.  For definiteness, we assume 1\,ks time lags at all temporal frequencies.  We have considered two average BAT-band flux levels ($F_{14-195\,{\rm keV}}=2\times 10^{-11}\ergpcmsqps$ and $F_{14-195\,{\rm keV}}=2\times 10^{-12}\ergpcmsqps$), two levels of variability (10\% and 20\% rms flux variability), and three observation lengths (100\,ks, 200\,ks and 400\,ks ``on-source'' exposure time).  For each pair of light curves, we add Poisson noise appropriate for the SXI, SXS and HXI.  We also introduce gaps in the lightcurve appropriate to a satellite in low Earth orbit.   We then conduct an analysis of the light curves in order to recover the time lags using the Fourier techniques laid out in \cite{zoghbi10}, treating the gaps with a maximum likelihood method
described in \cite{zoghbi13}.

We find that, for these parameters, the SXI and SXS cannot detect lags --- the count rates are simply insufficient.  However, lags in the Compton bump are readily detectable with the HXI.  Using simulated light curves in the 7--10\,keV and 20--50\,keV bands (i.e., a continuum dominated band and a band with a significant Compton-reflection contribution), Figure~\ref{fig:lags} shows the recovered lags as a function of Fourier frequency.  In the bright source scenario, we see that lags can be detected at an intermediate Fourier frequency in a 100\,ks observation, but a 200\,ks lightcurve is required in order to detect the lag across the Fourier spectrum.  This is important since, in real sources, the lags will change as a function of Fourier frequency, and these changes are crucial for disentangling reverberation delays from hard-lags that are intrinsic to the physics of the X-ray continuum emission.  In the faint source scenario, lags are detectable but a long (400\,ks) observation will be required to gain even crude information on the frequency dependence.

\section{Summary of Top Science}

We end this WP with a brief summary of the ``whys and hows'' of our top science goals.

\begin{enumerate}
\item {\bf OBJECTIVE : } Characterization of the ``torus'' and other high column density
obscuring structures in AGN. \\
{\bf WHY : }The basic nature of the torus is unclear; is it the reservoir for the accretion disk (inflow) or the outer parts of an AGN wind (outflow).  In many ways, the torus is the ``interface'' between the AGN and the rest of the galaxy, so understanding its nature is important for understanding the AGN-galaxy connection. We also see to understand the basic nature of CTAGN, poorly understood objects that are crucial in models of the CXB and SMBH evolution.\\
{\bf HOW : }We will use a combination of detailed iron line diagnostics (profile, shifts, Compton shoulder) enabled by the {\it ASTRO-H}/SXS as well as the broad-band/hard coverage of {\it ASTRO-H} to characterize transmission and reflection from the torus, even in cases where it may be Compton-Thick.

\item {\bf OBJECTIVE : } Define the X-ray continuum shape of AGN, from the softest energies observable to energies beyond their high-energy cutoff.\\
{\bf WHY : }A broad-band characterization of the continuum, correcting for absorption and reflection effects, is crucial for understanding the energetics of the AGN population and the true nature of the CXB.  Robust determinations of the high-energy cutoff are needed to access the physics of the corona.  \\
{\bf HOW : }Using the combination of all four instruments, {\it ASTRO-H} can cover the range 0.5--200\,keV with unprecedented sensitivity.  The SGD in particular will enable us to go beyond the exponentially-cutoff powerlaw approximation for the Comptonized high-energy spectrum.

\item {\bf OBJECTIVE : } Robust characterization of the inner accretion disk reflection even in
cases with complex absorption and composite soft excesses. \\
{\bf WHY} - We wish to probe accretion disk inclination, disk abundances, 
and black hole spin even in sources that are ``complicated'', i.e., have complex warm 
absorbers or highly variable spectral shapes.  Only then can we use disk reflection as a tool to 
undercover the true demographics of these properties. \\
{\bf HOW} - {\it ASTRO-H}/SXI+HXI will be the first observatory capable of, in a single observation, 
probing all three aspects of the disk reflection spectrum (the soft excess, broad iron line, and 
Compton hump).  SXS spectroscopy will be important for characterizing and allowing us to 
remove the effects of absorption.  Long SXI+HXI observations of bright and variable sources 
will allow us to see reverberation effects across the full spectrum.
\end{enumerate}

\clearpage
\begin{multicols}{2}
\end{multicols}

\end{document}